\documentclass{article}

\usepackage[utf8]{inputenc}
\usepackage{tismir}
\usepackage{amsmath}
\usepackage{hyperref}
\usepackage{url}
\usepackage{graphicx}
\usepackage{booktabs}
\usepackage{lipsum}
\usepackage{wasysym}

\usepackage{arydshln}

\DeclareMathOperator*{\argmax}{arg\,max}
\DeclareMathOperator*{\argmin}{arg\,min}

\title{Chord Recognition in Symbolic Music: A Segmental CRF Model, Segment-Level Features, and Comparative Evaluations on Classical and Popular Music}
\author{%
  Kristen Masada\thanks{School of Electrical Engineering and Computer Science, Ohio University, Athens, OH} ~and Razvan Bunescu\protect\footnotemark[1]}

\date{}

\authorref{Masada,~K. and Bunescu,~R.}
\authorshort{Kristen Masada and Razvan Bunescu} 
\titleshort{A Segmental CRF Model for Chord Recognition in Symbolic Music}

\begin{document}


\twocolumn[{%
\maketitleblock
\begin{abstract}
We present a new approach to harmonic analysis that is trained to segment music into a sequence of chord spans tagged with chord labels. Formulated as a semi-Markov Conditional Random Field (semi-CRF), this joint segmentation and labeling approach enables the use of a rich set of segment-level features, such as segment purity and chord coverage, that capture the extent to which the events in an entire segment of music are compatible with a candidate chord label. The new chord recognition model is evaluated extensively on three corpora of classical music and a newly created corpus of rock music. Experimental results show that the semi-CRF model performs substantially better than previous approaches when trained on a sufficient number of labeled examples and remains competitive when the amount of training data is limited.
\end{abstract}
}]
\saythanks{}


\section{Introduction and Motivation}
\label{sec:introduction}

Harmonic analysis is an important step towards creating high-level representations of tonal music. High-level structural relationships form an essential component of music analysis, whose aim is to achieve a deep understanding of how music works. At its most basic level, harmonic analysis of music in symbolic form requires the partitioning of a musical input into segments along the time dimension, such that the notes in each segment correspond to a musical chord. This {\it chord recognition} task can often be time consuming and cognitively demanding, hence the utility of computer-based implementations. Reflecting historical trends in artificial intelligence, automatic approaches to harmonic analysis have evolved from purely grammar-based and rule-based systems \citep{winograd:jmt68,maxwell:chapter92}, to systems employing weighted rules and optimization algorithms \citep{temperley:cmj99,pardo:cmj02,scholz:ismir08,rocher:icmc09}, to data driven approaches based on supervised machine learning (ML) \citep{raphael:ismir03,radicioni:amir10}. Due to their requirements for annotated data, ML approaches have also led to the development of music analysis datasets containing a large number of manually annotated harmonic structures, such as the 60 Bach chorales introduced in \citep{radicioni:amir10}, and the 27 themes and variations of TAVERN \citep{devaney:ismir15}.

\begin{figure}[t]
\centering
\includegraphics[width=\columnwidth]{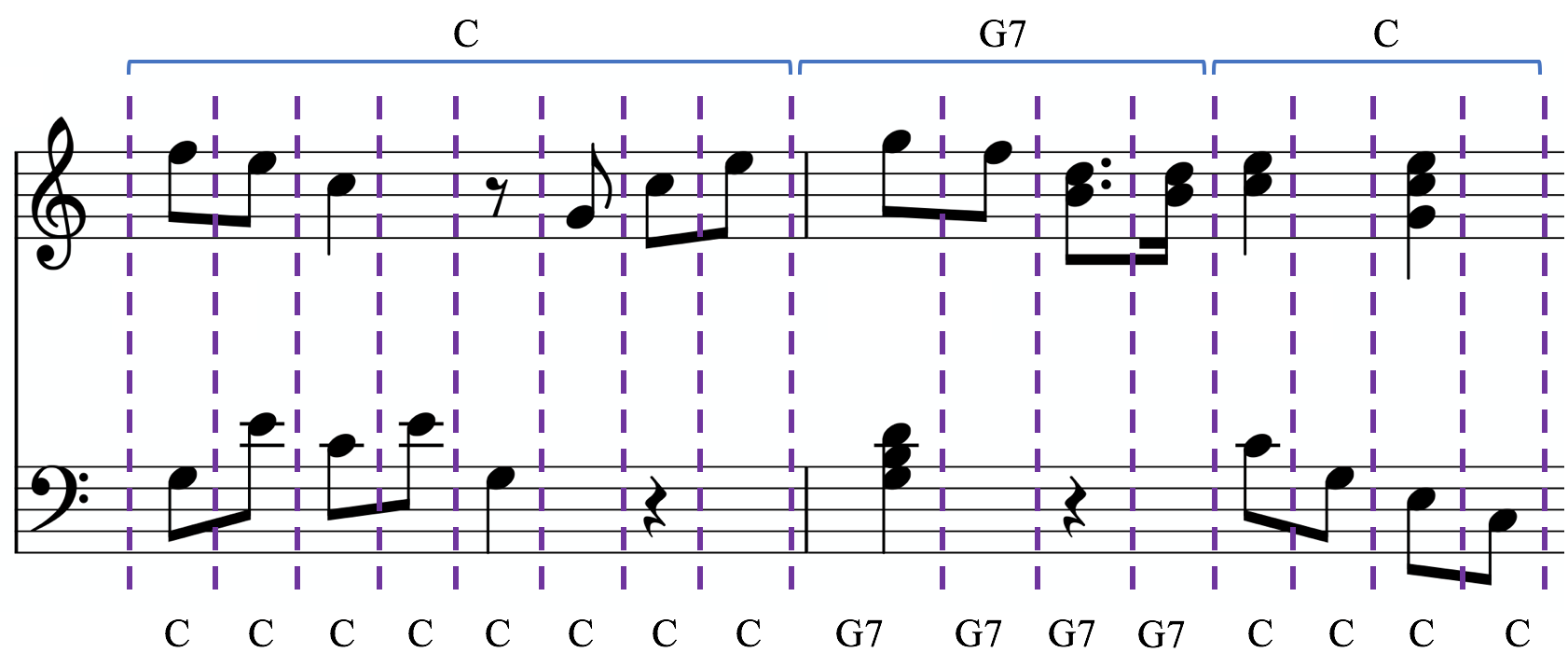}
\caption{Segment-based recognition (top) vs. event-based recognition (bottom) on measures 11 and 12 from Beethoven WoO68, using note onsets and offsets to create event boundaries.}
\label{fig:segments}
\end{figure}
In this work, we consider the music to be in symbolic form, i.e. as a collection of notes specified in terms of onset, offset, pitch, and metrical position. Symbolic representations can be extracted from formats such as MIDI, kern, or MusicXML. A relatively common strategy in ML approaches to chord recognition in symbolic music is to break the musical input into a sequence of short duration spans and then train sequence tagging algorithms such as Hidden Markov Models (HMMs) to assign a chord label to each span in the sequence (bottom of Figure~\ref{fig:segments}). The spans can result from quantization using a fixed musical period such as half a measure \citep{raphael:ismir03}. Alternatively, they can be constructed from consecutive note onsets and offsets \citep{radicioni:amir10}, as we also do in this paper. Variable-length chord segments are then created by joining consecutive spans labeled with the same chord symbol (at the top in Figure~\ref{fig:segments}).
A significant drawback of these short-span tagging approaches is that they do not explicitly model candidate segments during training and inference, consequently they cannot use segment-level features. Such features are needed, for example, to identify figuration notes (Appendix~\ref{sec:appendix-figuration}) or to help label segments that do not start with the root note. The chordal analysis system of \citet{pardo:cmj02} is an example where the assignment of chords to segments takes into account segment-based features, however the features have pre-defined weights and it uses a processing pipeline where segmentation is done independently of chord labeling.

In this paper, we propose a machine learning approach to chord recognition formulated under the framework of semi-Markov Conditional Random Fields (semi-CRFs). Also called segmental CRFs, this class of probabilistic graphical models can be trained to do joint segmentation and labeling of symbolic music (Section~\ref{sec:model}), using efficient Viterbi-based inference algorithms whose time complexity is linear in the length of the input. The system employs a set of chord labels (Section~\ref{sec:labels}) that correspond to the main types of tonal music chords (Appendix~\ref{sec:chords}) found in the evaluation datasets. Compared to HMMs and sequential CRFs which label the events in a sequence, segmental CRFs label candidate segments, as such they can exploit segment-level features. Correspondingly, we define a rich set of features that capture the extent to which the events in an entire segment of music are compatible with a candidate chord label (Section~\ref{sec:features}). The semi-CRF model incorporating these features is evaluated on three classical music datasets and a newly created dataset of popular music (Section~\ref{sec:datasets}). Experimental comparisons with two previous chord recognition models show that segmental CRFs obtain substantial improvements in performance on the three larger datasets, while also being competitive with the previous approaches on the smaller dataset (Section~\ref{sec:evaluation}).


\section{Semi-CRF Model for Chord Recognition}
\label{sec:model}

\begin{figure}[t]
\centering
\includegraphics[width=1.0\columnwidth]{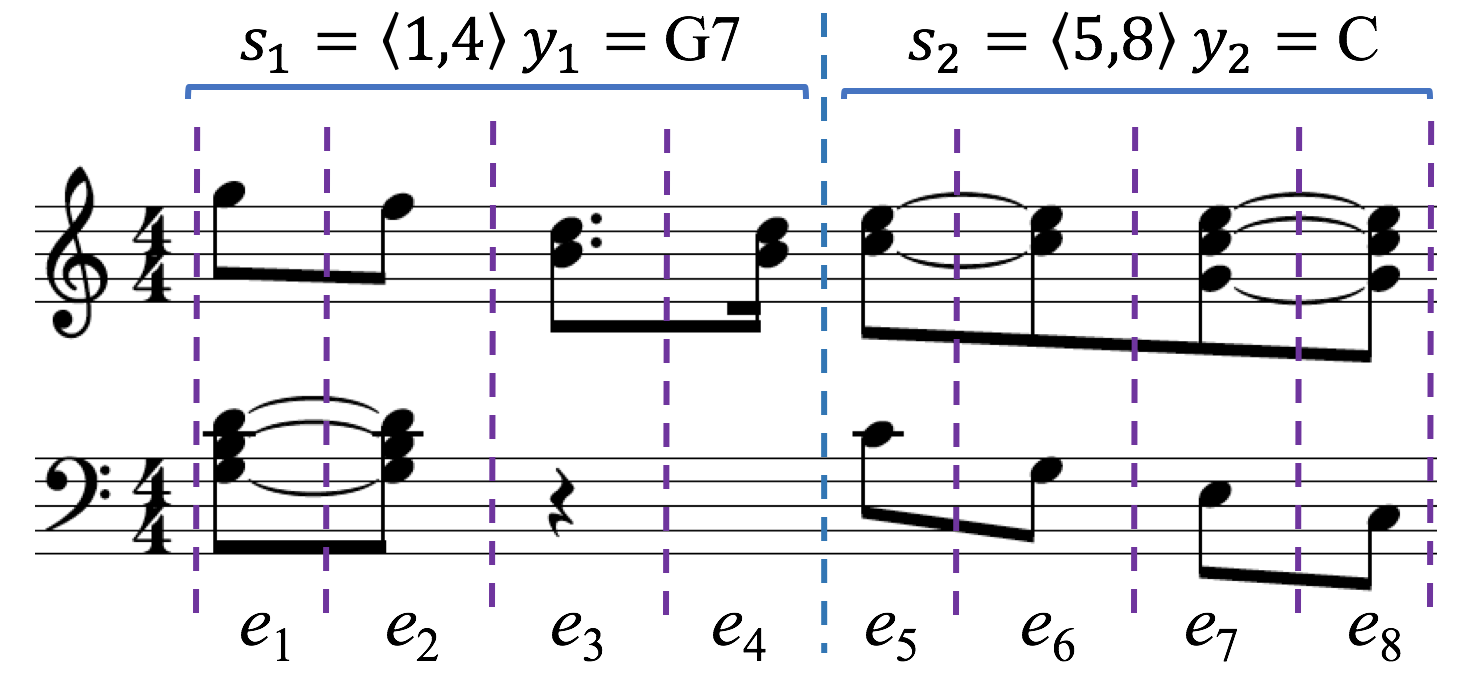}
\caption{Segment and labels (top) vs. events (bottom) for measure 12 from Beethoven WoO68.}
\label{fig:segments2}
\end{figure}

Since harmonic changes may occur only when notes begin or end, we first create a sorted list of all the note onsets and offsets in the input music, i.e. the list of {\it partition points} \citep{pardo:cmj02}, shown as vertical dotted lines in Figure~\ref{fig:segments}. A basic music {\it event} \citep{radicioni:amir10} is then defined as the set of pitches sounding in the time interval between two consecutive partition points. As an example, Table~\ref{tab:input-rep} provides the pitches and overall duration for each event shown in Figure~\ref{fig:segments2}. The segment number and chord label associated with each event are also included. Not shown in this table is a boolean value for each pitch indicating whether or not it is held over from the previous event. For instance, this value would be false for C5 and E5 appearing in event $e_5$, but true for C5 and E5 in event $e_6$.

\begin{table}[t]
\centering
  \begin{tabular}{lllll}
  \toprule
  Seg. & Label & Event & Pitches & Len. \\ \midrule
  $s_1$ & G7 & $e_1$ & G3, B3, D4, G5 & 1/8\\
   & G7 & $e_2$ & G3, B3, D4, F5 & 1/8\\
   & G7 & $e_3$ & B4, D5 & 3/16\\
   & G7 & $e_4$ & B4, D5 & 1/16\\ \hline 
  $s_2$ & C & $e_5$ & C4, C5, E5 & 1/8\\
   & C & $e_6$ & G3, C5, E5 & 1/8\\
   & C & $e_7$ & E3, G4, C5, E5 & 1/8\\
   & C & $e_8$ & C3, G4, C5, E5 & 1/8\\
  \bottomrule
  \end{tabular}
  \caption{Input representation for measure 12 from Beethoven WoO68, showing the pitches and duration for each event, as well as the corresponding segment and label, where G7 stands for G:maj:add7, and C stands for C:maj.}
\label{tab:input-rep}
\end{table}

Let $\mathbf{s} = \langle s_1, s_2, ..., s_K \rangle$ denote a segmentation of the musical input $\mathbf{x}$, where a segment $s_k = \langle s_k.f, s_k.l \rangle$ is identified by the positions $s_k.f$ and $s_k.l$ of its first and last events, respectively. Let $\mathbf{y} = \langle y_1, y_2, ..., y_K \rangle$ be the vector of chord labels corresponding to the segmentation $\mathbf{s}$. A semi-Markov CRF \citep{sarawagi:nips04} defines a probability distribution over segmentations and their labels as shown in Equations~\ref{eq:semi1} and \ref{eq:semi2}. Here, the global segmentation feature vector $\mathbf{F}$ decomposes as a sum of local segment feature vectors $\mathbf{f}(s_k, y_k, y_{k-1}, \mathbf{x})$, with label $y_0$ set to a constant ``no chord'' value. 
\begin{eqnarray}
  P(\mathbf{s}, \mathbf{y} | \mathbf{x}, \mathbf{w}) & = & \frac{e^{{\mathbf{w}}^T{\mathbf{F}(\mathbf{s}, \mathbf{y} ,\mathbf{x})}}}{Z(\mathbf{x})} \label{eq:semi1} \\
  \mathbf{F}(\mathbf{s}, \mathbf{y} ,\mathbf{x}) & = & \sum_{k = 1}^K \mathbf{f}(s_k, y_k, y_{k-1}, \mathbf{x})  \label{eq:semi2}
\end{eqnarray}
where $Z(\mathbf{x}) = \displaystyle \sum_{\mathbf{s}', \mathbf{y}'} e^{{\mathbf{w}}^T{\mathbf{F}(\mathbf{s}', \mathbf{y}' ,\mathbf{x})}}$ and $\mathbf{w}$ is a vector of parameters.

Following Muis and Lu \citep{muis:naacl16}, for faster inference, we further restrict the local segment features to two types: {\it segment-label features} $\mathbf{f}(s_k, y_k, \mathbf{x})$ that depend on the segment and its label, and label {\it transition features} $\mathbf{g}(y_k, y_{k-1}, \mathbf{x})$ that depend on the labels of the current and previous segments. The corresponding probability distribution over segmentations is shown in Equations~\ref{eq:weak1} to \ref{eq:weak3}, which use two vectors of parameters: $\mathbf{w}$ for segment-label features and $\mathbf{u}$ for transition features.
\begin{eqnarray}
  P(\mathbf{s}, \mathbf{y} | \mathbf{x}, \mathbf{w}, \mathbf{u}) \! & \! = \! & \! \frac{e^{{\mathbf{w}}^T{\mathbf{F}(\mathbf{s}, \mathbf{y} ,\mathbf{x})} + {\mathbf{u}}^T{\mathbf{G}(\mathbf{s}, \mathbf{y} ,\mathbf{x})}}}{Z(\mathbf{x})} \label{eq:weak1} \\
  \mathbf{F}(\mathbf{s}, \mathbf{y} ,\mathbf{x}) \! & \! = \! & \! \sum_{k = 1}^K \mathbf{f}(s_k, y_k, \mathbf{x})  \label{eq:weak2} \\
  \mathbf{G}(\mathbf{s}, \mathbf{y} ,\mathbf{x}) \! & \! = \! & \! \sum_{k = 1}^K \mathbf{g}(y_k, y_{k-1}, \mathbf{x})  \label{eq:weak3}
\end{eqnarray}

Given an arbitrary segment $s$ and a label $y$, the vector of segment-label features can be written as $\mathbf{f}(s, y, \mathbf{x}) = [f_1(s, y), ..., f_{|\mathbf{f}|}(s, y)]$, where the input $\mathbf{x}$ is left implicit in order to compress the notation. Similarly, given arbitrary labels $y$ and $y'$, the vector of label transition features can be written as  $\mathbf{g}(y, y', \mathbf{x}) = [g_1(y, y'), ..., g_{|\mathbf{g}|}(y, y')]$. In Section~\ref{sec:features} we describe the set of segment-label features $f_i(s, y)$ and label transition features $g_j(y, y')$ that are used in our semi-CRF chord recognition system.

\begin{figure}[h]
\centering
\includegraphics[width=0.6\columnwidth]{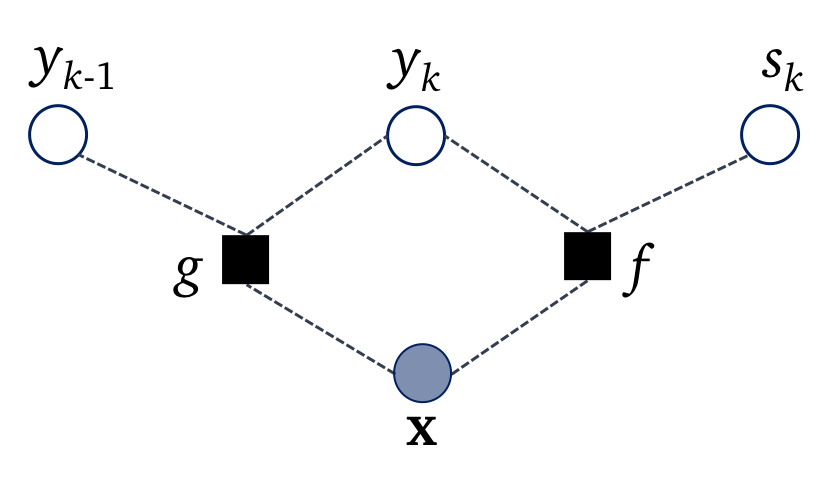}
\caption{Factor graph representation of the semi-CRF.}
\label{fig:fg}
\end{figure}

As probabilistic graphical models, semi-CRFs can be represented using factor graphs, as illustrated in Figure~\ref{fig:fg}. Factor graphs \citep{frey:ieee01} are bipartite graphs that express how a global function (e.g. $P(\mathbf{s}, \mathbf{y} | \mathbf{x}, \mathbf{w}, \mathbf{u})$) of many {\it variables} (e.g. $s_k$, $y_k$, and $\mathbf{x}$) factorizes into a product of local functions, or {\it factors}, (e.g. $f$ and $g$) defined over fewer variables.


Equations~\ref{eq:weak2} and~\ref{eq:weak3} show that the contribution of any given feature to the final log-likelihood score is given by summing up its value over all the segments (for local features $f$) or segment pairs (for local features $g$).
This design choice stems from two assumptions. First, we adopt the stationarity assumption, according to which the segment-label feature distribution does not change with the position in the music. Second, we use the Markov assumption, which implies that the label of a segment depends only on its boundaries and the labels of the adjacent segments. This assumption leads to the factorization of the probability distribution into a product of potentials.
Both the stationarity assumption and the Markov assumption are commonly used in ML models for structured outputs, such as linear CRFs \citep{lafferty:ml01}, semi-CRFs \citep{sarawagi:nips04}, HMMs \citep{rabiner:ieee89}, structural SVMs \citep{tsochantaridis:icml04}, or the structured perceptron \citep{collins:emnlp02} used in HMPerceptron. These assumptions lead to summing the same feature over multiple substructures in the overall output score, which makes {\it inference} and {\it learning} tractable using dynamic programming.

The {\it inference} problem for semi-CRFs refers to finding the most likely segmentation $\hat{\mathbf{s}}$ and its labeling $\hat{\mathbf{y}}$ for an input $\mathbf{x}$, given the model parameters. For the weak semi-CRF model in Equation~\ref{eq:weak1}, this corresponds to:
\begin{eqnarray}
   \!\!\!\!\!\!\!\!\hat{\mathbf{s}}, \hat{\mathbf{y}} \! & \!\!\!\! = \!\!\!\! & \! \argmax_{\mathbf{s}, \mathbf{y}} \; P(\mathbf{s}, \mathbf{y} | \mathbf{x}, \mathbf{w}, \mathbf{u}) \\
  \! & \!\!\!\! = \!\!\!\! & \! \argmax_{\mathbf{s}, \mathbf{y}} \; {\mathbf{w}}^T{\mathbf{F}(\mathbf{s}, \mathbf{y} ,\mathbf{x})} + {\mathbf{u}}^T{\mathbf{G}(\mathbf{s}, \mathbf{y} ,\mathbf{x})} \\
  \! & \!\!\!\! = \!\!\!\! & \! \argmax_{\mathbf{s}, \mathbf{y}} \; {\mathbf{w}}^T \!\! \sum_{k = 1}^K \mathbf{f}(s_k, y_k, \mathbf{x}) + {\mathbf{u}}^T \!\! \sum_{k = 1}^K \mathbf{g}(y_k, y_{k-1}, \mathbf{x}) \label{eq:optimization}
\end{eqnarray}
The maximum is taken over all possible labeled segmentations of the input, up to a maximum segment length. Correspondingly, $\mathbf{s}$ and $\mathbf{y}$ can be seen as ``candidate'' segmentations and ``candidate'' labelings, respectively. Their number is exponential in the length of the input, which rules out a brute-force search. However, due to the factorization into vectors of local features $f_i(s, y)$ and $g_j(y, y')$, it can be shown that the optimization problem from Equation~\ref{eq:optimization} can be solved with a semi-Markov analogue of the usual Viterbi algorithm. Let $L$ be a maximum segment length. Following \citep{sarawagi:nips04}, let $V(i, y)$ denote the largest value ${\mathbf{w}}^T{\mathbf{F}(\tilde{\mathbf{s}}, \tilde{\mathbf{y}}, \mathbf{x})} + {\mathbf{u}}^T{\mathbf{G}(\tilde{\mathbf{s}}, \tilde{\mathbf{y}} ,\mathbf{x})}$ of a partial segmentation $\tilde{\mathbf{s}}$ such that its last segment ends at position $i$ and has label $y$. Then $V(i, y)$ can be computed with the following dynamic programming recursion for $i = 1, 2, ..., |\mathbf{x}|$:
\begin{equation}
  V(i,\! y) \!=\!\! \displaystyle \max_{y', 1 \leq l \leq L} V(i\!-\!l,\! y') + {\mathbf{w}}^T \mathbf{f}(\langle i\!-\!l + 1, i \rangle, y, \mathbf{x}) + {\mathbf{u}}^T{\mathbf{g}(y, y',\mathbf{x})} 
\end{equation}
where the base cases are $V(0, y) = 0$ and $V(j, y) = -\infty$ if $j < 0$, and $\langle i - l + 1, i \rangle$ denotes the segment starting at position $i - l + 1$ and ending at position $i$.
Once $V(|\mathbf{x}|, y)$ is computed for all labels $y$, the best labeled segmentation can be recovered in linear time by following the path traced by $\max_y V(|\mathbf{x}|, y)$.

The {\it learning} problem for semi-CRFs refers to finding the model parameters that maximize the likelihood over a set of training sequences $T = \{\mathbf{x}_n, \mathbf{s}_n, \mathbf{y}_n\}_{n = 1}^N$. Usually this is done by minimizing the negative log-likelihood $-L(T;\mathbf{w}, \mathbf{u})$ and an L2 regularization term, as shown below for weak semi-CRFs:
\begin{equation}
  L(T; \mathbf{w}, \mathbf{u}) \! = \!\!  \sum_{n=1}^N {\mathbf{w}}^T{\mathbf{F}(\mathbf{s}_n, \mathbf{y}_n ,\mathbf{x}_n)} + {\mathbf{u}}^T{\mathbf{G}(\mathbf{s}_n, \mathbf{y}_n,\mathbf{x}_n)} - \log{Z(\mathbf{x}_n)}
\end{equation}
\begin{equation}
  \hat{\mathbf{w}}, \hat{\mathbf{u}} =  \argmin_{\mathbf{w}, \mathbf{u}} -L(T; \mathbf{w}, \mathbf{u}) + \frac{\lambda}{2}\left(||\mathbf{w}||^2 + ||\mathbf{u}||^2\right) 
\end{equation}
This is a convex optimization problem, which is solved with the L-BFGS procedure in the StatNLP package used to implement our system. The partition function $Z(\mathbf{x})$ and the feature expectations that appear in the gradient of the objective function are computed efficiently using a dynamic programming algorithm similar to the forward-backward procedure \citep{sarawagi:nips04}.

\section{Chord Recognition Labels}
\label{sec:labels}

A {\it chord} is a group of notes that form a cohesive harmonic unit to the listener when sounding simultaneously \citep{aldwell:book11}. As explained in Appendix~\ref{sec:chords}, we design our system to handle the following types of chords: {\it triads}, {\it augmented 6th chords}, {\it suspended chords}, and {\it power chords}. The chord labels used in previous chord recognition research range from coarse grained labels that indicate only the chord root \citep{temperley:cmj99} to fine grained labels that capture mode, inversions, added and missing notes \citep{harte:phd10}, and even chord function \citep{devaney:ismir15}. Here we follow the middle ground proposed by \citet{radicioni:amir10} and define a core set of labels for triads that encode the chord root (12 pitch classes), the mode (major, minor, diminished), and the added note (none, fourth, sixth, seventh), for a total of 144 different labels. For example, the label {\it C-major-none} for a simple C major triad corresponds to the combination of a root of {\it C} with a mode of {\it major} and no added note. This is different from the label {\it C-major-seventh} for a C major seventh chord, which corresponds to the combination of a root of {\it C} with a mode of {\it major} and an added note of {\it seventh}. Note that there is only one generic type of added seventh note, irrespective of whether the interval is a major, minor, or diminished seventh, which means that a C major seventh chord and a C dominant seventh chord are mapped to the same label. However, once the system recognizes a chord with an added seventh, determining whether it is a major, minor, or diminished seventh can be done accurately in a simple post-processing step: determine if the chord contains a non figuration note (defined in Appendix~\ref{sec:appendix-figuration}) that is 11, 10, or 9 half steps from the root, respectively, inverted or not, modulo 12. Once the type of the seventh interval is determined, it is straightforward to determine the type of seventh chord (dominant, major, minor, minor-major, fully diminished, or half-diminished) based on the mode of the chord (major, minor, or diminished).

Augmented sixth chords are modeled through a set of 36 labels that capture the lowest note (12 pitch classes) and the 3 types (Appendix~\ref{sec:aug6}). Similarly, suspended and power chords are modeled through a set of 48 labels that capture the root note (12 pitch classes) and the 4 types (Appendix~\ref{sec:suspow}).

Because the labels do not encode for function, the model does not require knowing the key in which the input was written. While the number of labels may seem large, the number of parameters in our model is largely independent of the number of labels. This is because we design the chord recognition features (Section~\ref{sec:features}) to not test for the chord root, which also enables the system to recognize chords that were not seen during training. The decision to not use the key context was partly motivated by the fact that 3 of the 4 datasets we used for experimental evaluation do not have functional annotations (see Section~\ref{sec:datasets}). Additionally, complete key annotation can be difficult to perform, both manually and automatically. Key changes occur gradually, thus making it difficult to determine the exact location where one key ends and another begins \citep{papadopoulos:dafx}. This makes locating modulations and tonicizations difficult and also hard to evaluate \citep{gomez:phd06}. At the same time, we recognize that harmonic analysis is not complete without functional analysis. Functional analysis features could also benefit the basic chord recognition task described in this paper. In particular, the chord transition features that we define in Appendix~\ref{sec:chord-bigrams} depend on the absolute distance in half steps between the roots of the chords. However, a V-I transition has a different distribution than a I-IV transition, even though the root distance is the same. Chord transition distributions also differ between minor and major keys. As such, using key context could further improve chord recognition.

\section{Chord Recognition Features}
\label{sec:features}
The semi-CRF model uses five major types of features, as described in detail in Appendix~\ref{sec:appendix-features}. Segment purity features compute the percentage of segment notes that belong to a given chord (Appendix~\ref{sec:purity}). We include these on the grounds that segments with a higher purity with respect to a chord are more likely to be labeled with that chord. Chord coverage features determine if each note in a given chord appears at least once in the segment (Appendix~\ref{sec:coverage}). Similar to segment purity, if the segment covers a higher percentage of the chord's notes, it is more likely to be labeled with that chord. Bass features determine which note of a given chord appears as the bass in the segment (Appendix~\ref{sec:bass}). For a correctly labeled segment, its bass note often matches the root of its chord label. If the bass note instead matches the chord's third or fifth, or is an added dissonance, this may indicate that the chord $y$ is inverted or incorrect. Chord bigram features capture chord transition information (Appendix~\ref{sec:chord-bigrams}). These features are useful in that the arrangement of chords in chord progressions is an important component of {\it harmonic syntax}. Finally, we include metrical accent features for chord changes, as chord segments are more likely to begin on accented beats (Appendix~\ref{sec:chord-accent}).

\section{Chord Recognition Datasets}
\label{sec:datasets}

For evaluation, we used four chord recognition datasets:
\begin{enumerate}
  \item BaCh: this is the Bach Choral Harmony Dataset, a corpus of 60 four-part Bach chorales that contains 5,664 events and 3,090 segments in total \citep{radicioni:amir10}.
  \item TAVERN: this is a corpus of 27 complete sets of themes and variations for piano, composed by Mozart and Beethoven. It consists of 63,876 events and 12,802 segments overall \citep{devaney:ismir15}.
  \item KP Corpus: the Kostka-Payne corpus is a dataset of 46 excerpts compiled by Bryan Pardo from Kostka and Payne's music theory textbook. It contains 3,888 events and 911 segments~\citep{kostka:book84}.
  \item Rock: this is a corpus of 59 pop and rock songs that we compiled from Hal Leonard's {\it The Best Rock Songs Ever (Easy Piano)} songbook. It is 25,621 events and 4,221 segments in length.
\end{enumerate}

\subsection{The Bach Chorale (BaCh) Dataset}

The BaCh corpus has been annotated by a human expert with chord labels, using the set of triad labels described in Section~\ref{sec:labels}. Of the 144 possible labels, 102 appear in the dataset and of these only 68 appear 5 times or more. Some of the chord labels used in the manual annotation are enharmonic, e.g. C-sharp major and D-flat major, or D-sharp major and E-flat major. Reliably producing one of two enharmonic chords cannot be expected from a system that is agnostic of the key context. Therefore, we normalize the chord labels and for each mode we define a set of 12 canonical roots, one for each scale degree. When two enharmonic chords are available for a given scale degree, we selected the one with the fewest sharps or flats in the corresponding key signature. Consequently, for the major mode we use the canonical root set \{C, Db, D, Eb, E, F, Gb, G, Ab, A, Bb, B\}, whereas for the minor and diminished modes we used the root set \{C, C\#, D, D\#, E, F, F\#, G, G\#, A, Bb, B\}. Thus, if a chord is manually labeled as C-sharp major, the label is automatically changed to the enharmonic D-flat major. The actual chord notes used in the music are left unchanged. Whether they are spelled with sharps or flats is immaterial, as long as they are enharmonic with the root, third, fifth, or added note of the labeled chord. After performing enharmonic normalization on the chords in the dataset, 90 labels remain.

\subsection{The TAVERN Dataset}

The TAVERN dataset\endnote{Link to TAVERN: \\ \url{https://github.com/jcdevaney/TAVERN}} currently contains 17 works by Beethoven (181 variations) and 10 by Mozart (100 variations). The themes and variations are divided into a total of 1,060 phrases, 939 in major and 121 in minor. The pieces have two levels of segmentations: chords and phrases. The chords are annotated with Roman numerals, using the Humdrum representation for functional harmony\endnote{Link to Humdrum: \\ \url{http://www.humdrum.org/Humdrum/representations/harm.rep.html}}. When finished, each phrase will have annotations from two different experts, with a third expert adjudicating cases of disagreement between the two. After adjudication, a unique annotation of each phrase is created and joined with the note data into a combined file encoded in standard **kern format. However, many pieces do not currently have the second annotation or the adjudicated version. Consequently, we only used the first annotation for each of the 27 sets. Furthermore, since our chord recognition approach is key agnostic, we developed a script that automatically translated the Roman numeral notation into the key-independent canonical set of labels used in BaCh. Because the TAVERN annotation does not mark added fourth notes, the only added chords that were generated by the translation script were those containing sixths and sevenths. This results in a set of 108 possible labels, of which 69 appear in the dataset.

\subsection{The Kostka and Payne Corpus}

The Kostka-Payne (KP) corpus\endnote{Link to Kostka-Payne corpus: \\ \url{http://www.cs.northwestern.edu/~pardo/kpcorpus.zip}} does not contain chords with added fourth or sixth notes. However, it includes fine-grained chord types that are outside of the label set of triads described in Section~\ref{sec:labels}, such as fully and half-diminished seventh chords, dominant seventh chords, and dominant seventh flat ninth chords. We map these seventh chord variants to the generic added seventh chords, as discussed in Section~\ref{sec:labels}. Chords with ninth intervals are mapped to the corresponding chord without the ninth in our label set. The KP Corpus also contains the three types of augmented 6th chords introduced in Appendix~\ref{sec:aug6}. 
Thus, by extending our chord set to include augmented 6th labels, there are 12 roots $\times$ 3 triad modes $\times$ 2 added notes + 12 bass notes $\times$ 3 aug6 modes = 108 possible labels overall. Of these, 76 appear in the dataset.

A number of MIDI files in the KP corpus contain unlabeled sections at the beginning of the song. These sections also appear as unlabeled in the original Kostka-Payne textbook. We omitted these sections from our evaluation, and also did not include them in the KP Corpus event and segment counts. Bryan Pardo's original MIDI files for the KP Corpus also contain several missing chords, as well as chord labels that are shifted from their true onsets. We used chord and beat list files sent to us by David Temperley to correct these mistakes.

\subsection{The Rock Dataset}
\label{sec:rock-dataset}

To evaluate the system's ability to recognize chords in a different genre, we compiled a corpus of 59 pop and rock songs from Hal Leonard's {\it The Best Rock Songs Ever (Easy Piano)} songbook. Like the KP Corpus, the Rock dataset contains chords with added ninths---including major ninth chords and dominant seventh chords with a sharpened ninth---as well as inverted chords. We omit the ninth and inversion numbers in these cases. Unique from the other datasets, the Rock dataset also possesses suspended and power chords. We extend our chord set to include these, adding suspended second, suspended fourth, dominant seventh suspended fourth, and power chords. We use the major mode canonical root set for suspended second and power chords and the minor canonical root set for suspended fourth chords, as this configuration produces the least number of accidentals. In all, there are 12 roots $\times$ 3 triad modes $\times$ 4 added notes + 12 roots $\times$ 4 sus and pow modes = 192 possible labels, with only 48 appearing in the dataset.

\def\arraystretch{1.2}
\begin{table*}[t]
\small
\centering
  \begin{tabular}{l r r c | c c | c c | c c | c c | c c |}
  \cline{5-14}
   & & & & \multicolumn{4}{|c}{Full chord evaluation} &  \multicolumn{6}{|c|}{Root-level evaluation}\\ \cline{2-14}
   & \multicolumn{3}{|c|}{Statistics} & \multicolumn{2}{c|}{semi-CRF} & \multicolumn{2}{c|}{HMPerceptron} & \multicolumn{2}{c|}{semi-CRF} & \multicolumn{2}{c|}{HMPerceptron} & \multicolumn{2}{c|}{Melisma} \\ \hline
   \multicolumn{1}{|l|}{Dataset} & Events & Seg.'s & Labels & Acc$_E$ & F$_S$ &  Acc$_E$ & F$_S$ &  Acc$_E$ & F$_S$ &  Acc$_E$ & F$_S$ &  Acc$_E$ & F$_S$\\ \hline
   \multicolumn{1}{|l|}{BaCh} & 5,664 & 3,090 & 90 & {\bf 83.2} & {\bf 77.5} & 77.2 & 69.9 & {\bf 88.9 } & {\bf 84.2 } & 84.8 & 77.0 & 84.3 & 74.7\\
   \multicolumn{1}{|l|}{TAVERN} & 63,876 & 12,802 & 69 & {\bf 78.0} & {\bf 64.0} & 57.0 & 22.5 & {\bf 86.0} & {\bf 71.4} & 69.2 & 33.2 & 76.7 & 41.5\\
   \multicolumn{1}{|l|}{KPCorpus} & 3,888 & 911 & 76 & {\bf 73.0} & {\bf 53.0} & 72.9 & 45.4 & 79.3 & 59.0 & 79.0 & 51.9 & {\bf 81.9} & {\bf 62.2}\\
   \multicolumn{1}{|l|}{Rock} & 25,621 & 4,221 & 48 & {\bf 70.1} & {\bf 55.9} & 61.3 & 34.6 & {\bf 86.1} & {\bf 65.1} & 80.7 & 42.9 & 77.9 & 36.3\\
  \hline
  \end{tabular}
  \caption{Dataset statistics and summary of results (event-level accuracy Acc$_E$ and segment-level F-measure F$_S$).}
\label{tab:sum-chord}
\end{table*}
\def\arraystretch{1}

Similar to the KP Corpus, unlabeled segments occur at the beginning of some songs, which we omit from evaluation. Additionally, the Rock dataset uses an N.C. (i.e. no chord) label for some segments within songs where the chord is unclear. We broke songs containing this label into subsections consisting of the segments occurring before and after each N.C. segment, discarding subsections less than three measures long. 

To create the Rock dataset, we converted printed sheet music to MusicXML files using the optical music recognition (OMR) software PhotoScore\endnote{Link to PhotoScore: \\ \url{http://www.neuratron.com/photoscore.htm}}. We noticed in the process of making the dataset that some of the originally annotated labels were incorrect. For instance, some segments with added note labels were missing the added note, while other segments were missing the root or were labeled with an incorrect mode. We automatically detected these cases and corrected each label by hand, considering context and genre-specific theory. We also omitted two songs (`Takin' Care of Business' and `I Love Rock N' Roll') from the 61 songs in the original Hal Leonard songbook, the former because of its atonality and the latter because of a high percentage of mistakes in the original labels.

\section{Experimental Evaluation}
\label{sec:evaluation}

We implemented the semi-Markov CRF chord recognition system using a multi-threaded package\endnote{Link to StatNLP: \\ \url{http://statnlp.org/research/ie/}} that has been previously used for noun-phrase chunking of informal text \citep{muis:naacl16}. The following sections describe the experimental results obtained on the four datasets from Section~\ref{sec:datasets} for: our semi-CRF system; Radicioni and Esposito's perceptron-trained HMM system, HMPerceptron; and Temperley's computational music system, Melisma Music Analyzer\endnote{Link to David Temperley's Melisma Music Analyzer: \\ \url{http://www.link.cs.cmu.edu/melisma/}}.
When interpretting these results, it is important to consider a number of important differences among the three systems:
\begin{itemize}
  \item HMPerceptron and semi-CRF are data driven, therefore their performance depends on the number of training examples available. Both approaches are agnostic of music theoretic principles such as harmony changing primarily on strong metric positions, however they can learn such tendencies to the extent they are present in the training data.
  \item Compared to HMPerceptron, semi-CRFs can use segment-level features. Besides this conceptual difference, the semi-CRF system described here uses a much larger number of features than the HMPerceptron system, which by itself can lead to better performance but may also require more training examples.
  \item Both Melisma and HMPerceptron use metrical accents automatically induced by Melisma, whereas semi-CRF uses the Music21 accents derived from the notated meter. The more accurate notated meter could favor the semi-CRF system, although results in Section~\ref{sec:bach-eval} show that, at least on BaCh, HMPerceptron does not benefit from using the notated meter.
\end{itemize}

Table~\ref{tab:sum-chord} shows a summary of the full chord and root-level experimental results provided in this section. Two overall types of measures are used to evaluate a system's performance on a dataset: event-level accuracy (Acc$_E$) and segment-level F-measure (F$_S$). Acc$_E$ simply refers to the percentage of events for which the system predicts the correct label out of the total number of events in the dataset. Segment-level F-measure is computed based on precision and recall, two evaluation measures commonly used in information retrieval \citep{baeza-yates:book99}, as follows:
\begin{itemize}
  \item Precision (P$_S$) is the percentage of segments predicted correctly by the system out of the total number of segments that it predicts (correctly or incorrectly) for all songs in the dataset.
  \item Recall (R$_S$) is the percentage of segments predicted correctly out of the total number of segments annotated in the original score for all songs in the dataset.
  \item F-Measure (F$_S$) is the harmonic mean between P$_S$ and R$_S$, i.e. $F_S = 2 P_S R_S / (P_S + R_S)$.
\end{itemize}
Note that a predicted segment is considered correct if and only if both its boundaries and its label match those of a true segment.

\subsection{BaCh Evaluation}
\label{sec:bach-eval}

We evaluated the semi-CRF model on BaCh using 10-fold cross validation: the 60 Bach chorales were randomly split into a set of 10 folds, and each fold was used as test data, with the other nine folds being used for training. We then evaluated HMPerceptron using the same randomly generated folds to enable comparison with our system. However, we noticed that the performance of HMPerceptron could vary significantly between two different random partitions of the data into folds. Therefore, we repeated the 10-fold cross validation experiment 10 times, each time shuffling the 60 Bach chorales and partitioning them into 10 folds. For each experiment, the test results from the 10 folds were pooled together  and one value was computed for each performance measure (accuracy, precision, recall, and F-measure). The overall performance measures for the two systems were then computed by averaging over the 10 values (one from each experiment). The sample standard deviation for each performance measure was also computed over the same 10 values.

For semi-CRF, we computed the frequency of occurrence of each feature in the training data, using only the true segment boundaries and their labels. To speedup training and reduce overfitting, we only used features whose counts were at least 5. The performance measures were computed by averaging the results from the 10 test folds for each of the fold sets. Table~\ref{tab:bach-chord} shows the averaged event-level and segment-level performance of the semi-CRF model, together with two versions of the HMPerceptron: HMPerceptron$_1$, for which we do enharmonic normalization both on training and test data, similar to the normalization done for semi-CRF; and HMPerceptron$_2$, which is the original system from \citep{radicioni:amir10} that does enharmonic normalization only on test data.
\begin{table}[!h]
\centering
  \begin{tabular}{lllll}
  \multicolumn{5}{c}{BaCh: Full chord evaluation}\\
  \toprule
  System & Acc$_E$ & P$_S$ & R$_S$ & F$_S$ \\ \midrule
  semi-CRF & {\bf 83.2} & {\bf 79.4} & {\bf 75.8} & {\bf 77.5}\\[-4pt]
           & \multicolumn{1}{r}{\scriptsize 0.2} & \multicolumn{1}{r}{\scriptsize 0.2} & \multicolumn{1}{r}{\scriptsize 0.2} & \multicolumn{1}{r}{\scriptsize 0.2} \\
  HMPerceptron$_1$ & 77.2 & 71.2 & 68.8 & 69.9 \\[-4pt]
           & \multicolumn{1}{r}{\scriptsize 2.1} & \multicolumn{1}{r}{\scriptsize 2.0} & \multicolumn{1}{r}{\scriptsize 2.2} & \multicolumn{1}{r}{\scriptsize 1.8} \\
  HMPerceptron$_2$ & 77.0 & 71.0 & 68.5 & 69.7 \\[-4pt]
           & \multicolumn{1}{r}{\scriptsize 2.1} & \multicolumn{1}{r}{\scriptsize 2.0} & \multicolumn{1}{r}{\scriptsize 2.3} & \multicolumn{1}{r}{\scriptsize 1.8} \\
  \bottomrule
  \end{tabular}
  \caption{Comparative results (\%) and standard deviations on the BaCh dataset, using Event-level {\bf accuracy } (Acc$_E$) and Segment-level {\bf precision} (P$_S$), {\bf recall }(R$_S$), and {\bf F-measure } (F$_S$).}
\label{tab:bach-chord}
\end{table}

The semi-CRF model achieves a 6.2\% improvement in event-level accuracy over the original model HMPerceptron$_2$, which corresponds to a 27.0\% relative error reduction\footnote{$27\% = (83.2 - 77.0)/(100 - 77.0)$}. The improvement in accuracy over HMPerceptron$_1$ is statistically significant at an averaged $p$-value of 0.001, using a one-tailed Welch's t-test over the sample of 60 chorale results for each of the 10 fold sets. The improvement in segment-level performance is even more substantial, with a 7.8\% absolute improvement in F-measure over the original HMPerceptron$_2$ model, and a 7.6\% improvement in F-measure over the HMPerceptron$_1$ version, which is statistically significant at an averaged $p$-value of 0.002, using a one-tailed Welch's t-test. The standard deviation values computed for both event-level accuracy and F-Measure are about one order of magnitude smaller for semi-CRF than for HMPerceptron, demonstrating that the semi-CRF is also more stable than the HMPerceptron. As HMPerceptron$_1$ outperforms HMPerceptron$_2$ in both event and segment-level accuracies, we will use HMPerceptron$_1$ for the remaining evaluations and will simply refer to it as HMPerceptron.

\begin{table}[!h]
\centering
  \begin{tabular}{lllll}
  \multicolumn{5}{c}{BaCh: Root only evaluation}\\
  \toprule
  System & Acc$_E$ & P$_S$ & R$_S$ & F$_S$ \\ \midrule
  semi-CRF & {\bf 88.9} & {\bf 85.4} & {\bf 83.0} & {\bf 84.2}\\
  HMPerceptron & 84.8 & 78.0 & 76.2 & 77.0 \\
  Melisma & 84.3 & 73.2 & 76.3 & 74.7 \\
  \bottomrule
  \end{tabular}
  \caption{Root only results (\%) on the BaCh dataset, using Event-level {\bf accuracy } (Acc$_E$) and Segment-level {\bf precision } (P$_S$), {\bf recall } (R$_S$), and {\bf F-measure } (F$_S$).}
\label{tab:bach-root}
\end{table}

We also evaluated performance in terms of predicting the correct root of the chord, e.g. if the true chord label were C:maj, a predicted chord of C:maj:add7 would still be considered correct, because it has the same root as the correct label. We performed this evaluation for semi-CRF, HMPerceptron, and the harmony component of Temperley's Melisma. Results show that semi-CRF improves upon the event-level accuracy of HMPerceptron by 4.1\%, producing a relative error reduction of 27.0\%, and that of Melisma by 4.6\%. Semi-CRF also achieves an F-measure that is 7.2\% higher than HMPerceptron and 9.5\% higher than Melisma. These improvements are statistically significant with a $p$-value of 0.01 using a one-tailed Welch's t-test.

\begin{table}[!h]
\centering
  \begin{tabular}{lllll}
  \multicolumn{5}{c}{BaCh: Metrical accent evaluation of semi-CRF}\\
  \toprule
  System & Acc$_E$ & P$_S$ & R$_S$ & F$_S$ \\ \midrule
  With accent & {\bf 83.6} & {\bf 79.6} & {\bf 75.9} & {\bf 77.6}\\
  Without accent & 77.7 & 74.8 & 68.0 & 71.2\\
  \bottomrule
  \end{tabular}
  \caption{Full chord Event (Acc$_E$) and Segment-level (P$_S$, R$_S$, F$_S$) results (\%) on the BaCh dataset, with and without metrical accent features.}
\label{tab:bach-accent}
\end{table}
Metrical accent is important for harmonic analysis: chord changes tend to happen in strong metrical positions; figuration such as passing and neighboring tones appear in metrically weak positions, whereas suspensions appear on metrically strong beats. We verified empirically the importance of metrical accent by evaluating the semi-CRF model on a random fold set from the BaCh corpus with and without all accent-based features. The results from Table~\ref{tab:bach-accent} show a substantial decrease in accuracy when the accent-based features are removed from the system.

Finally, we ran an evaluation of HMPerceptron on a random fold set from BaCh in two scenarios: HMPerceptron with Melisma metrical accent and HMPerceptron with Music21 accent. The results did not show a significant difference: with Melisma accent the event accuracy was 79.8\% for an F-measure of 70.2\%, whereas with Music21 accent the event accuracy was 79.8\% for an F-measure of 70.3\%. This negligible difference is likely due to the fact that HMPerceptron uses only coarse-grained accent information, i.e. whether a position is accented (Melisma accent 3 or more) or not accented (Melisma accent less than 3).

\subsubsection{BaCh Error Analysis}
Error analysis revealed wrong predictions being made on chords that contained dissonances that spanned the duration of the entire segment (e.g. a second above the root of the annotated chord), likely due to an insufficient number of such examples during training. Manual inspection also revealed a non-trivial number of cases in which we disagreed with the manually annotated chords, e.g. some chord labels were clear mistakes, as they did not contain any of the notes in the chord. This further illustrates the necessity of building music analysis datasets that are annotated by multiple experts, with adjudication steps akin to the ones followed by TAVERN.

\subsection{TAVERN Evaluation}
\label{sec:tavern-eval}
To evaluate on the TAVERN corpus, we created a fixed training-test split: 6 Beethoven sets ({\it B063}, {\it B064}, {\it B065}, {\it B066}, {\it B068}, {\it B069}) and 4 Mozart sets ({\it K025}, {\it K179}, {\it K265}, {\it K353}) were used for testing, while the remaining 11 Beethoven sets and 6 Mozart sets were used for training. All sets were normalized enharmonically before being used for training or testing. Table~\ref{tab:tavern-chord} shows the event-level and segment-level performance of the semi-CRF and HMPerceptron model on the TAVERN dataset.
\begin{table}[!h]
\centering
  \begin{tabular}{lllll}
  \multicolumn{5}{c}{TAVERN: Full chord evaluation}\\
  \toprule
  System & Acc$_E$ & P$_S$ & R$_S$ & F$_S$ \\ \midrule
  semi-CRF & {\bf 78.0} & {\bf 67.3} & {\bf 60.9} & {\bf 64.0} \\
  HMPerceptron & 57.0 & 24.5 & 20.8 & 22.5\\
  \bottomrule
  \end{tabular}
  \caption{Event (Acc$_E$) and Segment-level (P$_S$, R$_S$, F$_S$) results (\%) on the TAVERN dataset.}
\label{tab:tavern-chord}
\end{table}

As shown in Table~\ref{tab:tavern-chord}, semi-CRF outperforms HMPerceptron by 21.0\% for event-level chord evaluation and by 41.5\% in terms of chord-level F-measure. Root only evaluations provided in Table~\ref{tab:tavern-root} reveal that semi-CRF improves upon HMPerceptron's event-level root accuracy by 16.8\% and Melisma's event accuracy by 9.3\%. Semi-CRF also produces a segment-level F-measure value that is 38.2\% higher than that of HMPerceptron and 29.9\% higher than that of Melisma. These improvements are statistically significant with a $p$-value of 0.01 using a one-tailed Welch's t-test.

\begin{table}[!h]
\centering
  \begin{tabular}{lllll}
  \multicolumn{5}{c}{TAVERN: Root only evaluation}\\
  \toprule
  System & Acc$_E$ & P$_S$ & R$_S$ & F$_S$ \\ \midrule
  semi-CRF & {\bf 86.0} & {\bf 74.6} & {\bf 68.4} & {\bf 71.4} \\
  HMPerceptron & 69.2 & 38.2 & 29.4 & 33.2 \\
  Melisma & 76.7 & 42.3 & 40.7 & 41.5 \\
  \bottomrule
  \end{tabular}
  \caption{Event (Acc$_E$) and Segment-level (P$_S$, R$_S$, F$_S$) results (\%) on the TAVERN dataset.}
\label{tab:tavern-root}
\end{table}

\subsubsection{TAVERN Error Analysis}
\begin{figure}[t]
\centering
\includegraphics[width=0.8\columnwidth]{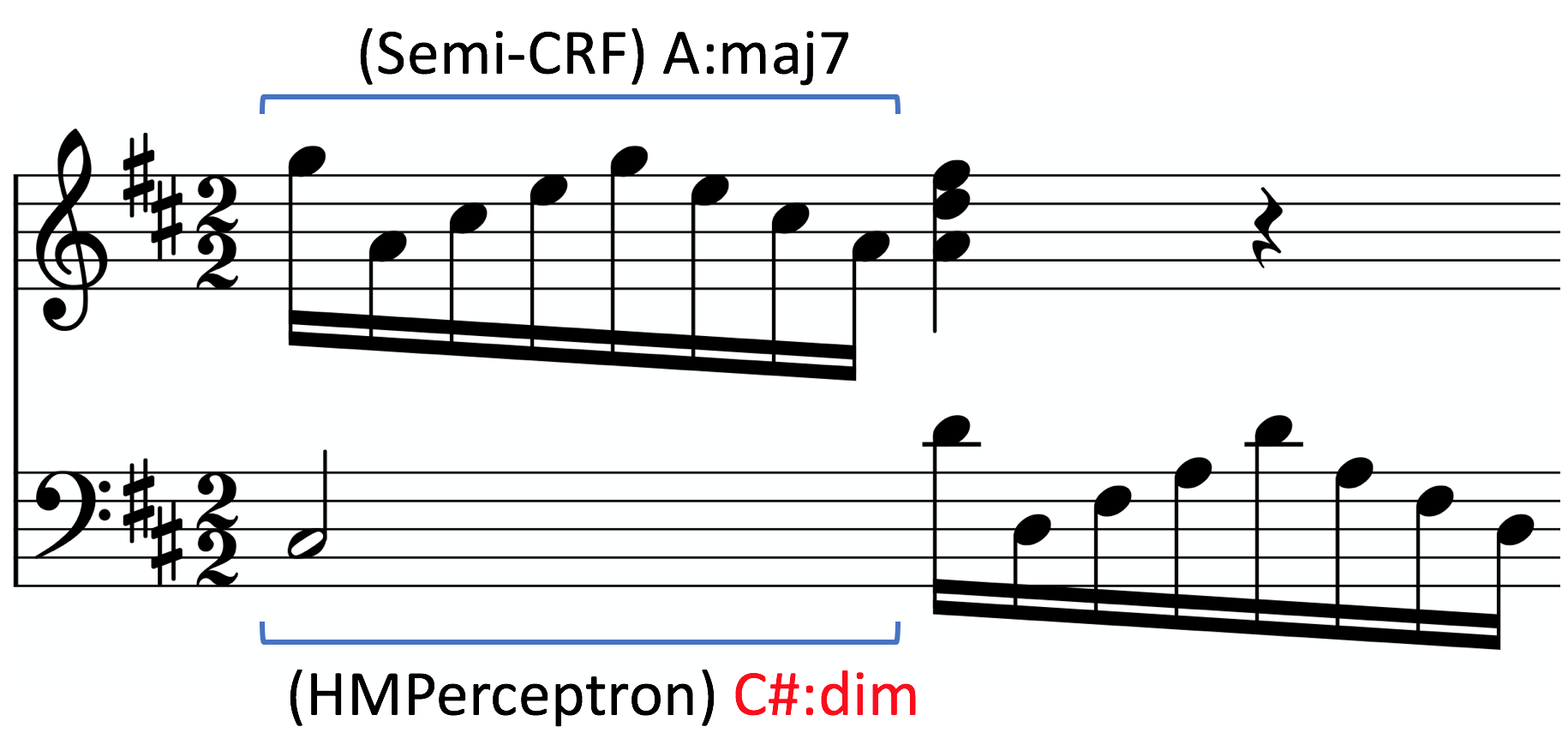}
\caption{Semi-CRF correctly predicts A:maj7 (top) for the first beat of measure 55 from Mozart K025, while HMPtron predicts C\#:dim (bottom).}
\label{fig:error-analysis}
\end{figure}

\begin{figure}[t]
\centering
\includegraphics[width=1.0\columnwidth]{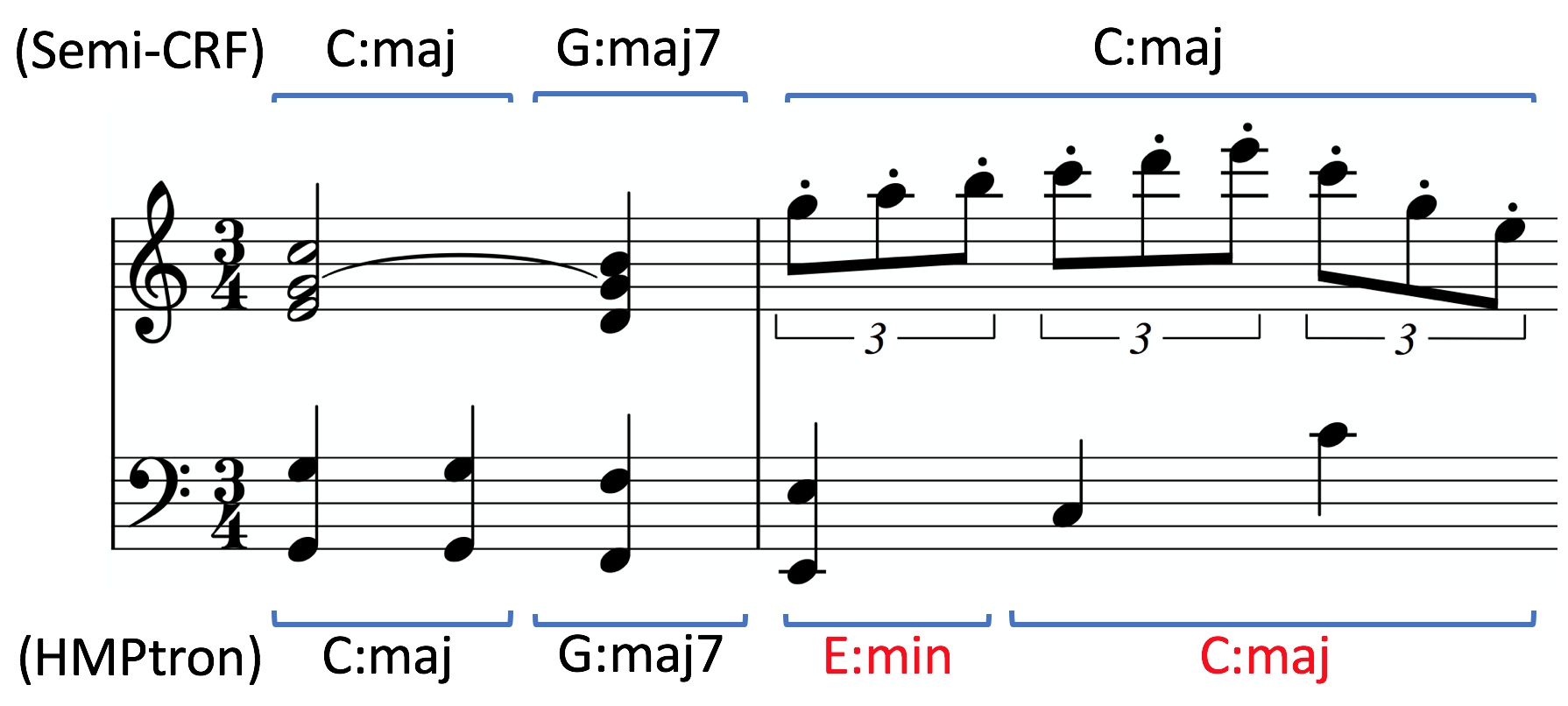}
\caption{Semi-CRF correctly predicts C:maj (top) for all of measure 280 from Mozart K179, while HMPtron predicts E:min (bottom) for the first beat and C:maj for the other two beats (bottom).}
\label{fig:error-analysis2}
\end{figure}
The results in Tables~\ref{tab:bach-chord} and~\ref{tab:tavern-chord} show that chord recognition is substantially more difficult in the TAVERN dataset than in BaCh. The comparatively lower performance on TAVERN is likely due to the substantially larger number of figurations and higher rhythmic diversity of the variations compared to the easier, mostly note-for-note texture of the chorales. Error analysis on TAVERN revealed many segments where the first event did not contain the root of the chord, such as in Figures~\ref{fig:error-analysis} and~\ref{fig:error-analysis2}. For such segments, HMPerceptron incorrectly assigned chord labels whose root matched the bass of this first event. Since a single wrongly labeled event invalidates the entire segment, this can explain the larger discrepancy between the event-level accuracy and the segment-level performance. In contrast, semi-CRF assigned the correct labels in these cases, likely due to its ability to exploit context through segment-level features, such as the chord root coverage feature $f_4$ and its duration-weighted version $f_{11}$. In the case of Figure~\ref{fig:error-analysis}, C\# appears in the bass of the first beat of the measure and HMPerceptron incorrectly predicts a segment with label C\#:dim for this beat. In contrast, semi-CRF correctly predicts the label A:maj7 for this segment. In Figure~\ref{fig:error-analysis2}, semi-CRF correctly predicts a C:maj segment that lasts for the entirety of the measure, while HMPerceptron predicts an E:min segment for the first beat, as E appears doubled in the bass here.

\subsection{KP Corpus Evaluation}
\label{sec:kpcorpus-eval}

To evaluate on the full KP Corpus dataset, we split the songs into 11 folds. In this configuration, 9 folds contain 4 songs each, while the remaining 2 folds contain 5 songs. We then created two versions of semi-CRF: the original system without augmented 6th chord features (semi-CRF$_1$) and a system with augmented 6th features (semi-CRF$_2$). We tested both versions on all 46 songs, as shown in Table~\ref{tab:kpcorpus1-chord}. We could not perform the same evaluation on HMPerceptron because it was not designed to handle augmented 6th chords.
\begin{table}[!h]
\centering
  \begin{tabular}{lllll}
  \multicolumn{5}{c}{KP Corpus 46 songs: Full chord evaluation}\\
  \toprule
  System & Acc$_E$ & P$_S$ & R$_S$ & F$_S$ \\ \midrule
  semi-CRF$_1$ & 72.0 & 59.0 & 49.2 & 53.5\\
  semi-CRF$_2$ & {\bf 73.4} & {\bf 59.6} & {\bf 50.1} & {\bf 54.3} \\
  \bottomrule
  \end{tabular}
  \caption{Event (Acc$_E$) and Segment-level (P$_S$, R$_S$, F$_S$) results (\%) on the KP Corpus dataset.}
\label{tab:kpcorpus1-chord}
\end{table}

The results in Table~\ref{tab:kpcorpus1-chord} demonstrate the utility of adding augmented 6th chord features to our system, as semi-CRF$_2$ outperforms semi-CRF$_1$ on all measures. 
We will use semi-CRF$_2$ for the rest of the evaluations in this section, simply calling it semi-CRF.

We additionally perform root only evaluation on the full dataset for semi-CRF and Melisma. We ignore events that belong to the true augmented 6th chord segments when computing the root accuracies for both systems, as augmented 6th chords technically do not contain a root note. As shown in Table~\ref{tab:kpcorpus1-root}, Melisma is only marginally better than semi-CRF in terms of event-level root accuracy, however it has a segment-level F-measure that is 1.1\% better.
\begin{table}[!h]
\centering
  \begin{tabular}{lllll}
  \multicolumn{5}{c}{KP Corpus 46 songs: Root only evaluation}\\
  \toprule
  System & Acc$_E$ & P$_S$ & R$_S$ & F$_S$ \\ \midrule
  semi-CRF & 80.7 & {\bf 66.3} & 56.2 & 60.8 \\
  Melisma & {\bf 80.9} & 60.6 & {\bf 63.3 } & {\bf 61.9} \\
  \bottomrule
  \end{tabular}
  \caption{Event (Acc$_E$) and Segment-level (P$_S$, R$_S$, F$_S$) results (\%) on the KP Corpus dataset.}
\label{tab:kpcorpus1-root}
\end{table}

To enable comparison with HMPerceptron, we also evaluate all systems on the 36 songs that do not contain augmented 6th chords. Because of the reduced number of songs available for training, we used leave-one-out evaluation for both semi-CRF and HMPerceptron. Table~\ref{tab:kpcorpus2-chord} shows that semi-CRF obtains a marginal improvement in chord event accuracy and a more substantial 7.6\% improvement in segment-level F-measure in comparison with HMPerceptron. The comparative results in Table~\ref{tab:kpcorpus2-root} show that Melisma outperforms both machine learning systems for root only evaluation. Nevertheless, the semi-CRF is still competitive with Melisma in terms of both event-level accuracy and segment-level F-measure.
\begin{table}[!h]
\centering
  \begin{tabular}{lllll}
  \multicolumn{5}{c}{KP Corpus 36 songs: Full chord evaluation}\\
  \toprule
  System & Acc$_E$ & P$_S$ & R$_S$ & F$_S$ \\ \midrule
  semi-CRF & {\bf 73.0} & {\bf 55.6} & {\bf 50.7} & {\bf 53.0} \\
  HMPerceptron & 72.9 & 48.2 & 43.6 & 45.4\\
  \bottomrule
  \end{tabular}
  \caption{Event (Acc$_E$) and Segment-level (P$_S$, R$_S$, F$_S$) results (\%) on the KP Corpus dataset.}
\label{tab:kpcorpus2-chord}
\end{table}

\begin{table}[!h]
\centering
  \begin{tabular}{lllll}
  \multicolumn{5}{c}{KP Corpus 36 songs: Root only evaluation}\\
  \toprule
  System & Acc$_E$ & P$_S$ & R$_S$ & F$_S$ \\ \midrule
  semi-CRF & 79.3 & {\bf 61.8} & 56.4 & 59.0 \\
  HMPerceptron & 79.0 & 54.7 & 49.9 & 51.9\\
  Melisma & {\bf 81.9 } & 60.7 & {\bf 63.7} & {\bf 62.2} \\
  \bottomrule
  \end{tabular}
  \caption{Event (Acc$_E$) and Segment-level (P$_S$, R$_S$, F$_S$) results (\%) on the KP Corpus dataset.}
\label{tab:kpcorpus2-root}
\end{table}

We additionally compare semi-CRF against the HarmAn algorithm created by \citet{pardo:cmj02}, which achieves a 75.8\% event-level accuracy on the KP Corpus. We made several modifications to the initial evaluation of semi-CRF on the full KP Corpus to enable this comparison. For instance, Pardo and Birmingham omit a Schumann piece from their evaluation, testing HarmAn on 45 songs instead of 46. We omitted this piece as well. They also look at the labels that appear in the dataset beforehand, ignoring any segments whose correct labels are chords that appear less than 2\% of the time when rounded. We followed suit with this, ignoring segments labeled with augmented 6th chords and other less common labels. Overall, semi-CRF obtains an event-level accuracy of 75.3\%, demonstrating that it is competitive with HarmAn. However, it is important to note that these results are still not fully comparable: sometimes HarmAn predicts multiple labels for a single segment, and when the correct label is among these, Pardo and Birmingham divide by the number of labels the system predicts and consider this fractional value to be correct. In contrast, semi-CRF always predicts one label per segment.

\subsubsection{KP Corpus Error Analysis}

Both machine learning systems struggled on the KP corpus, with Melisma performing better on both event-level accuracy and segment-level F-measure. This can be explained by the smaller dataset, and thus the smaller number of available training examples. The KP corpus was the smallest of the four datasets, especially in terms of the number of segments -- less than a third compared to BaCh, and less than a tenth compared to TAVERN. Furthermore, the textbook excerpts are more diverse, as they are taken from 11 composers and are meant to illustrate a wide variety of music theory concepts, leading to mismatch between the training and test distributions and thus lower test performance.

\subsection{Rock Evaluation}
\label{sec:rock-eval}

We split the 59 songs in the rock dataset into 10 folds: 9 folds with 6 songs and 1 fold with 5 songs. Similar to the full KP Corpus evaluation from Section~\ref{sec:kpcorpus-eval}, we create two versions of the semi-CRF model. The first is the original semi-CRF system (semi-CRF$_1$) which does not contain suspended and power chord features. The second is a new version of semi-CRF (semi-CRF$_3$) which has suspended and power chord features added to it.
We do not include HMPerceptron in the evaluation of the full dataset, as it is not designed for suspended and power chords.
\begin{table}[!h]
\centering
  \begin{tabular}{lllll}
  \multicolumn{5}{c}{Rock 59 songs: Full chord evaluation}\\
  \toprule
  System & Acc$_E$ & P$_S$ & R$_S$ & F$_S$ \\ \midrule
  semi-CRF$_1$ &  66.0 & 49.8 & 47.3 & 48.5 \\
  semi-CRF$_3$ & {\bf 69.4} & {\bf 62.0} & {\bf 54.9} & {\bf 58.3} \\
  \bottomrule
  \end{tabular}
  \caption{Event (Acc$_E$) and Segment-level (P$_S$, R$_S$, F$_S$) results (\%) on the Rock dataset.}
\label{tab:rock1-chord}
\end{table}

As shown in Table~\ref{tab:rock1-chord}, semi-CRF$_3$ obtains higher event and segment-level accuracies than semi-CRF$_1$. Therefore, we use semi-CRF$_3$ for the rest of the experiments, simply calling it semi-CRF.

We perform root only evaluation on the full Rock dataset using semi-CRF and Melisma. In this case, it is not necessary to omit the true segments whose labels are suspended or power chords, as these types of chords contain a root. As shown in Table~\ref{tab:rock1-root}, semi-CRF outperforms Melisma on all measures: it obtains a 8.4\% improvement in event-level root accuracy and a 31.5\% improvement in segment-level F-measure over Melisma.

\begin{table}[!h]
\centering
  \begin{tabular}{lllll}
  \multicolumn{5}{c}{Rock 59 songs: Root only evaluation}\\
  \toprule
  System & Acc$_E$ & P$_S$ & R$_S$ & F$_S$ \\ \midrule
  semi-CRF & {\bf 85.8} & {\bf 70.9} & {\bf 63.2} & {\bf 66.8} \\
  Melisma & 77.4 & 29.5 & 44.0 & 35.3 \\
  \bottomrule
  \end{tabular}
  \caption{Event (Acc$_E$) and Segment-level (P$_S$, R$_S$, F$_S$) results (\%) on the Rock dataset.}
\label{tab:rock1-root}
\end{table}

We also evaluate only on the 51 songs that do not contain suspended or power chords to compare semi-CRF against HMPerceptron. We do this by splitting the reduced number of songs into 10 folds: 9 folds with 5 test songs and 46 training songs, and 1 fold with 6 test songs and 45 training songs. The results shown in Table~\ref{tab:rock2-chord} demonstrate that semi-CRF performs better than HMPerceptron: it achieves an 8.8\% improvement in event-level chord accuracy and a 21.3\% improvement in F-measure over HMPerceptron. Additionally, we evaluate the root-level performance of all systems on the 51 songs. The results in Table~\ref{tab:rock2-root} show that the semi-CRF achieves better root-level accuracy than both systems: it obtains a 5.4\% improvement in event-level root accuracy over HMPerceptron and a 8.2\% improvement over Melisma. In terms of segment-level accuracy, it demonstrates a 22.2\% improvement in F-measure over HMPerceptron and a 28.8\% improvement over Melisma. These results are statistically significant with a $p$-value of 0.01 using a one-tailed Welch's t-test.
\begin{table}[!h]
\centering
  \begin{tabular}{lllll}
  \multicolumn{5}{c}{Rock 51 songs: Full chord evaluation}\\
  \toprule
  System & Acc$_E$ & P$_S$ & R$_S$ & F$_S$ \\ \midrule
  semi-CRF & {\bf 70.1} & {\bf 58.8} & {\bf 53.2} & {\bf 55.9} \\
  HMPerceptron & 61.3 & 41.0 & 29.9 & 34.6\\
  \bottomrule
  \end{tabular}
  \caption{Event (Acc$_E$) and Segment-level (P$_S$, R$_S$, F$_S$) results (\%) on the Rock dataset.}
\label{tab:rock2-chord}
\end{table}

\begin{table}[!h]
\centering
  \begin{tabular}{lllll}
  \multicolumn{5}{c}{Rock 51 songs: Root only evaluation}\\
  \toprule
  System & Acc$_E$ & P$_S$ & R$_S$ & F$_S$ \\ \midrule
  semi-CRF & {\bf 86.1} & {\bf 68.6} & {\bf 61.9} & {\bf 65.1} \\
  HMPerceptron & 80.7 & 51.3 & 36.9 & 42.9\\
  Melisma & 77.9 & 30.6 & 45.8 & 36.3 \\
  \bottomrule
  \end{tabular}
  \caption{Event (Acc$_E$) and Segment-level (P$_S$, R$_S$, F$_S$) results (\%) on the Rock dataset.}
\label{tab:rock2-root}
\end{table}

 \subsubsection{Rock Error Analysis}
 
As mentioned in Section~\ref{sec:rock-dataset}, we automatically detected and manually fixed a number of mistakes that we found in the original chord annotations. In some instances, although the root of the provided chord label was missing from the corresponding segment, the label was in fact correct. In these instances, it was often the case that the root appeared in the previous segment and thus was still perceptually salient to the listener, either because of its long duration or because it appeared in the last event of the previous segment. Sometimes, the same harmonic and melodic patterns were repeated throughout the piece, with the root appearing in the first few repetitions of these patterns, but disappearing later on. This was true for `Twist and Shout' by the Beatles, in which the same I IV V7 progression of C major, F major, and G dominant 7 is repeated throughout the song, with the root C disappearing from C major segments by measure 11. Due to their inability to exploit larger scale patterns, neither system could predict the correct label for such segments.
 
We also found that three of the songs that we manually detected as having labels with incorrect modes (`Great Balls of Fire,' `Heartbreak Hotel,' and `Shake, Rattle, and Roll') were heavily influenced by blues. The three songs contain many major chord segments where the major third is purposefully swapped for a minor third to create a blues feel. We kept the labels as they were in these instances, but again both systems struggled to correctly predict the true label in these cases.

\begin{figure}[t]
\centering
\includegraphics[width=0.65\columnwidth]{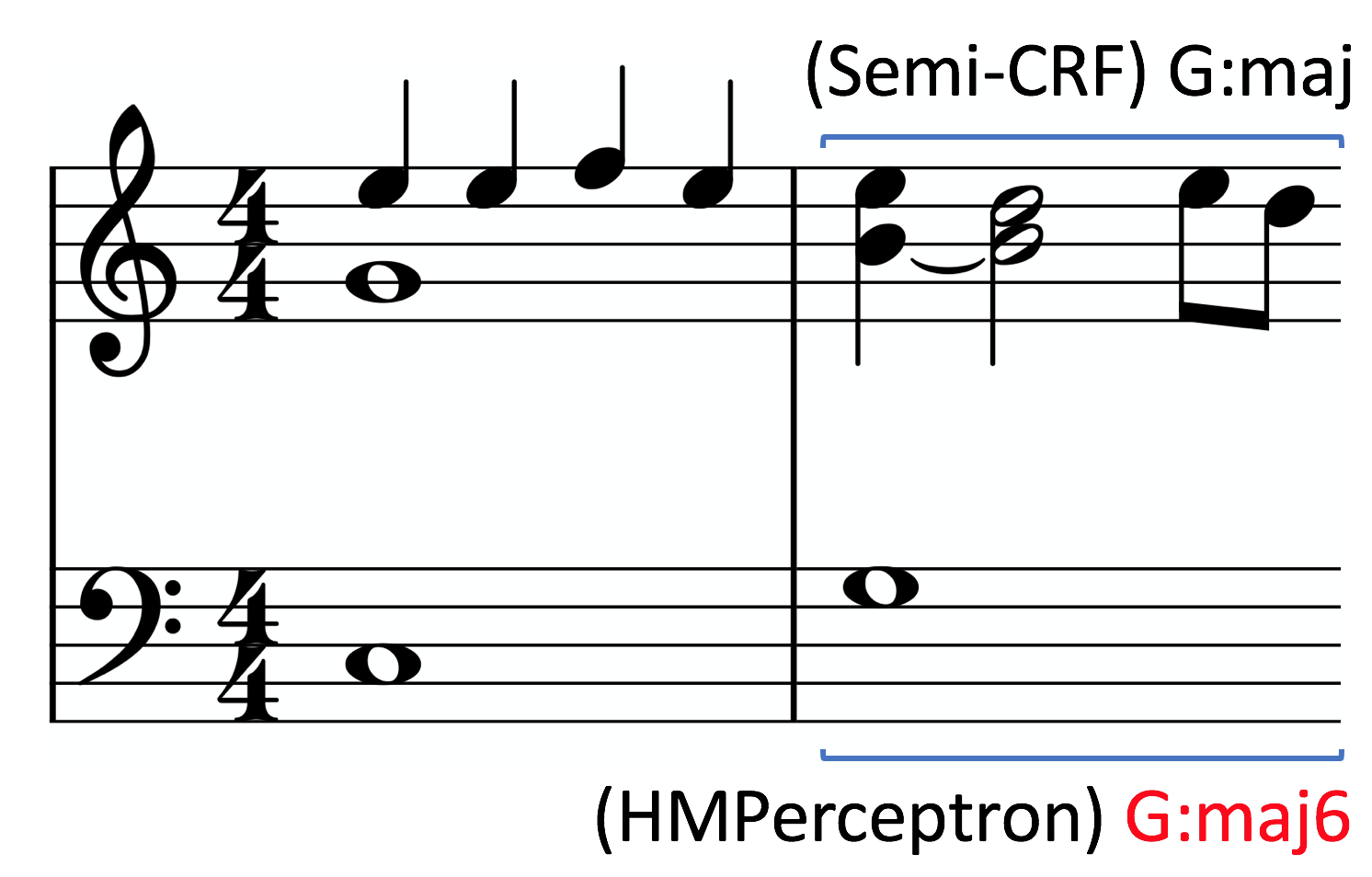}
\caption{Measures 14-15 of `Let It Be' by the Beatles, where HMPerceptron incorrectly predicts G:maj6 for measure 15 (bottom), while semi-CRF correctly predicts G:maj (top).}
\label{fig:error_analysis3}
\end{figure}

Figure~\ref{fig:error_analysis3} contains a brief excerpt from `Let It Be' by the Beatles demonstrating the utility of a segmental approach over an event-based approach. Semi-CRF correctly predicts a segment spanning measure 15 with the label G:maj, while HMPerceptron predicts these same segment boundaries, but incorrectly produces the label G:maj:add6. Semi-CRF most likely predicts the correct label because of its ability to heuristically detect figuration: the E5 on the first beat of measure 15 is a suspension, while the E5 on the fourth beat is a neighboring tone. It would be difficult for an event-based approach to recognize these notes as nonharmonic tones, as detecting figuration requires segment information. For instance, to detect a neighbor, this requires determining if one of its anchor notes belongs to the candidate segment (see Appendix~\ref{sec:appendix-figuration} for a full definition of neighbor and anchor tones).

\section{Related Work}

Numerous approaches for computerized harmonic analysis have been proposed over the years, starting with the pioneering system of \citet{winograd:jmt68}, in which a systemic grammar was used to encode knowledge of harmony. \citet{barthelemy:ismir01} and more recently \citet{rizo:chapter16} provide a good survey of previous work in harmonic analysis of symbolic music. Here, we focus on the three systems that inspired our work: Melisma \citep{temperley:cmj99}, HarmAn \citep{pardo:cmj02}, and HMPerceptron \citep{radicioni:amir10} (listed in chronological order). These systems, as well as our semi-CRF approach, incorporate knowledge of music theory through manually defined {\it rules} or {\it features}. For example, the ``compatibility rule'' used in Melisma is analogous to the chord coverage features used in the semi-CRF, the ``positive evidence'' score computed based on the six template classes in HarmAn, or the ``Asserted-notes'' features in HMPerceptron. Likewise, the segment purity features used in semi-CRF are analogous to the ``negative evidence'' scores from HarmAn, while the figuration heuristics used in semi-CRF can be seen as the counterpart of the ``ornamental dissonance rule'' used in Melisma. In these systems, each rule or feature is assigned an importance, or weight, in order to enable the calculation of an overall score for any candidate chord segmentation. Given a set of weights, optimization algorithms are used to determine the maximum scoring segmentation and labeling of the musical input. HMPerceptron uses the Viterbi algorithm \citep{rabiner:ieee89} to find the optimal sequence of event labels, whereas semi-CRF uses a generalization of Viterbi \citep{sarawagi:nips04} to find the joint most likely segmentation and labeling. The dynamic programming algorithm used in Melisma is actually an instantiation of the same general Viterbi algorithm -- like HMPerceptron and semi-CRF it makes a first-order Markov assumption and computes a similar lattice structure that enables a linear time complexity in the length of the input. HarmAn, on the other hand, uses the Relaxation algorithm \citep{cormen:book09}, whose original quadratic complexity is reduced to linear through a greedy approximation.

While the four systems are similar in terms of the musical knowledge they incorporate and their optimization algorithms, there are two important aspects that differentiate them:
\begin{enumerate}
  \item Are the weights learned from the data, or pre-specified by an expert? HMPerceptron and semi-CRF train their parameters, whereas Melisma and HarmAn have parameters that are predefined manually.
  \item Is chord recognition done as a joint segmentation and labeling of the input, or as a labeling of event sequences? HarmAn and semi-CRF are in the segment-based labeling category, whereas Melisma and HMPerceptron are event-based.
\end{enumerate}
Learning the weights from the data is more feasible, more scalable, and, given a sufficient amount of training examples, much more likely to lead to optimal performance. Furthermore, the segment-level classification has the advantage of enabling segment-level features that can be more informative than event-level analogues. The semi-CRF approach described in this paper is the first to take advantage of both learning the weights and performing a joint segmentation and labeling of the input.

\section{Future Work}
\label{sec:conclusion}

Manually engineering features for chord recognition is a cognitively demanding and time consuming process that requires music theoretical knowledge and that is not guaranteed to lead to optimal performance, especially when complex features are required. In future work we plan to investigate automatic feature extraction using recurrent neural networks (RNN). While RNNs can theoretically learn useful features from raw musical input, they are still event-level taggers, even when used in more sophisticated configurations, such as bi-directional deep LSTMs \citep{graves:book12}. We plan to use the Segmental RNNs of \citet{kong:iclr16}, which combine the benefits of RNNs and semi-CRFs: bidirectional RNNs compute representations of candidate segments, whereas segment-label compatibility scores are integrated using a semi-Markov CRF. Learning the features entirely from scratch could require a larger number of training examples, which may not be feasible to obtain. An alternative is to combine RNN sequence models with explicit knowledge of music theory, as was done recently by \cite{jaques:icml17} for the task of melody generation.

Music analysis tasks are mutually dependent on each other. Voice separation and chord recognition, for example, have interdependencies, such as figuration notes belonging to the same voice as their anchor notes. \citet{temperley:cmj99} note that harmonic analysis, in particular chord changes, can benefit meter modeling, whereas knowledge of meter is deemed crucial for chord recognition. This ``serious chicken-and-egg problem'' can be addressed by modeling the interdependent tasks together, for which probabilistic graphical models are a natural choice. Correspondingly, we plan to develop models that jointly solve multiple music analysis tasks, an approach that reflects more closely the way humans process music.


\section{Conclusion}

We presented a semi-Markov CRF model that approaches chord recognition as a joint segmentation and labeling task. Compared to event-level tagging approaches based on HMMs or linear CRFs, the segment-level approach has the advantage that it can accommodate features that consider all the notes in a candidate segment. This capability was shown to be especially useful for music with complex textures that diverge from the simpler note-for-note structures of the Bach chorales. The semi-CRF's parameters are trained on music annotated with chord labels, a data-driven approach that is more feasible than manually tuning the parameters, especially when the number of rules or features is large. Empirical evaluations on three datasets of classical music and a newly created dataset of rock music show that the semi-CRF model performs substantially better than previous approaches when trained on a sufficient number of labeled examples and stays competitive when the training data is small. The code is made publicly available on the first author's GitHub\endnote{Link to Code: \\ \url{https://github.com/kristenmasada/chord_recognition_semi_crf}}.


\theendnotes{}

\section*{Acknowledgments}

We would like to thank Patrick Gray for his help with pre-processing the TAVERN corpus. We thank Daniele P. Radicioni and Roberto Esposito for their help and their willingness to share the original BaCh dataset and the HMPerceptron implementation. We would like to thank Bryan Pardo for sharing the KP Corpus and David Temperley for providing us with an updated version that fixed some of the labeling errors in the original KP corpus. We would also like to thank David Chelberg for providing valuable comments on a previous version of the manuscript. Finally, we would like to thank the anonymous reviewers and the editorial team for their insightful comments.

Kristen Masada's work on this project was partly supported by an undergraduate research apprenticeship grant from the Honors Tutorial College at Ohio University.



\bibliography{tismir18}

\appendix

\section{Types of Chords in Tonal Music}
\label{sec:chords}

A {\it chord} is a group of notes that form a cohesive harmonic unit to the listener when sounding simultaneously \citep{aldwell:book11}. We design our system to handle the following types of chords: triads, augmented 6th chords, suspended chords, and power chords.

\subsection{Triads}
\label{sec:triads}

\begin{figure}[h]
\centering
\includegraphics[width=\columnwidth]{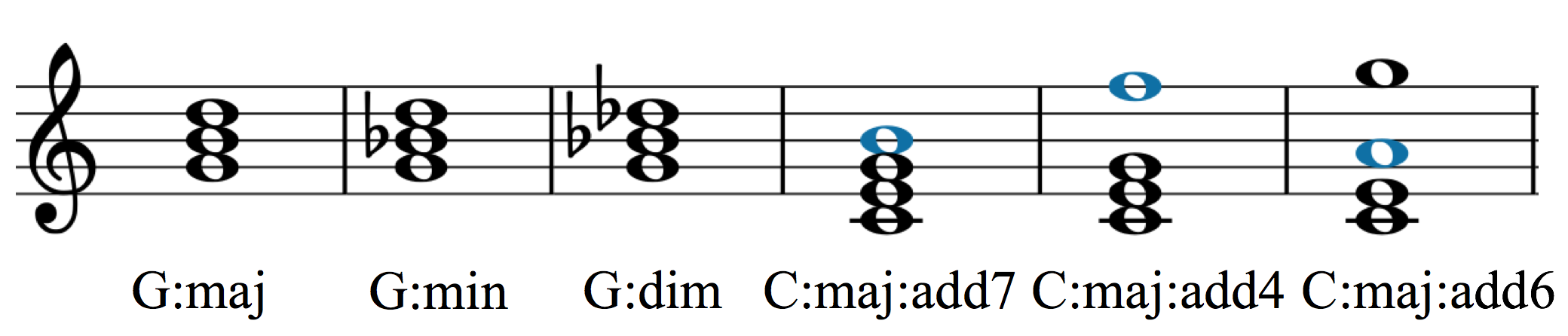}
\caption{Triads in 3 modes and with 3 added notes.}
\label{fig:chords-1}
\end{figure}

A {\it triad} is the prototypical instance of a chord. It is based on a root note, which forms the lowest note of a chord in standard position. A third and a fifth are then built on top of this root to create a three-note chord. Inverted triads also exist, where the third or fifth instead appears as the lowest note. The chord labels used in our system do not distinguish among inversions of the same chord. However, once the basic triad is determined by the system, finding its inversion can be done in a straightforward post-processing step, as a function of the bass note in the chord. The quality of the third and fifth intervals of a chord in standard position determines the mode of a triad. For our system, we consider three triad modes: {\it major} (maj), {\it minor} (min), and {\it diminished} (dim). A major triad consists of a major third interval (i.e. 4 half steps) between the root and third, as well as a perfect fifth (7 half steps) between the root and fifth. A minor triad has a minor third interval (3 half steps) between the root and third. Lastly, a diminished triad maintains the minor third between the root and third, but contains a diminished fifth (6 half steps) between the root and fifth. Figure~\ref{fig:chords-1} shows these three triad modes, together with three versions of a C major chord, one for each possible type of added note, as explained below.

A triad can contain an added note, or a fourth note. We include three possible added notes in our system: a fourth, a sixth, and a seventh. A fourth chord (add4) contains an interval of a perfect fourth (5 half steps) between the root and the added note for all modes. In contrast, the interval between the root and added note of a sixth chord (add6) of any mode is a major sixth (9 half steps). For seventh chords (add7), the added note interval varies. If the triad is major, the added note can form a major seventh (11 half steps) with the root, called a major seventh chord. It can also form a minor seventh (10 half steps) to create a dominant seventh chord. If the triad is minor, the added seventh can again either form an interval of a major seventh, creating a minor-major seventh chord, or a minor seventh, forming a minor seventh chord. Finally, diminished triads most frequently contain a diminished seventh interval (9 half steps), producing a fully diminished seventh chord, or a minor seventh interval, creating a half-diminished seventh chord.

\subsection{Augmented 6th Chords}
\label{sec:aug6}

\begin{figure}[h]
\centering
\includegraphics[width=0.65\columnwidth]{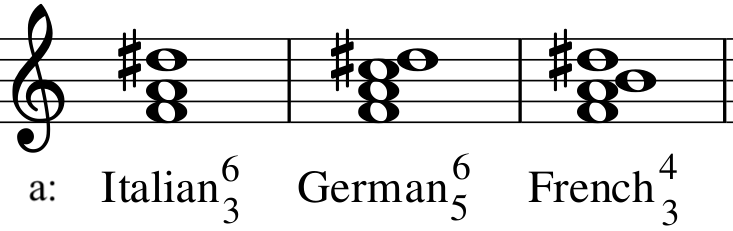}
\caption{Common types of augmented 6th chords, shown for the A minor scale. The same notes would also be used for the A major scale.}
\label{fig:chords-2}
\end{figure}

An {\it augmented 6th chord} is a type of chromatic chord defined by an augmented sixth interval between the lowest and highest notes of the chord \citep{aldwell:book11}. The three most common types of augmented 6th chords are {\it Italian}, {\it German}, and {\it French} sixth chords, as shown in Figure~\ref{fig:chords-2} in the key of A minor. In a minor scale, Italian sixth chords can be seen as iv chords with a sharpened root, in the first inversion. Thus, they can be created by stacking the sixth, first, and sharpened fourth scale degrees. In minor, German sixth chords are iv\textsuperscript{7} (i.e. minor seventh) chords with a sharpened root, in the first inversion. They are formed by combining the sixth, first, third, and sharpened fourth scale degrees. Lastly, French sixth chords are created by stacking the sixth, first, second, and sharpened fourth scale degrees. Thus, they are ii\textsuperscript{\diameter7} (i.e. half-diminished seventh) chords with a sharpened third, in second inversion.

\subsection{Suspended and Power Chords}
\label{sec:suspow}

\begin{figure}[h]
\centering
\includegraphics[width=0.75\columnwidth]{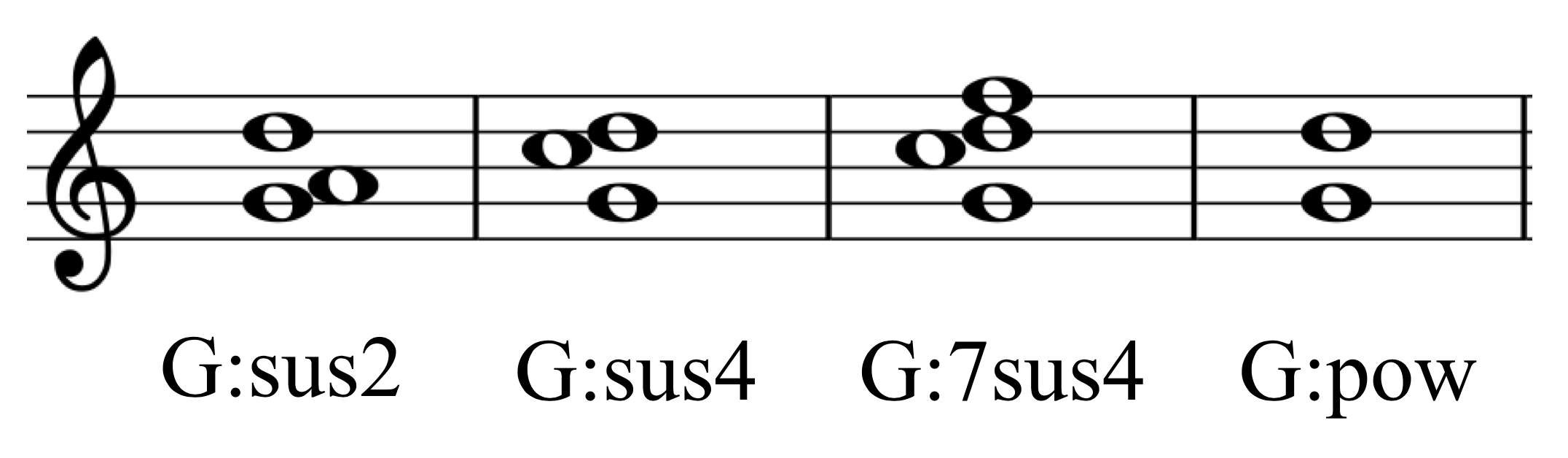}
\caption{Suspended and power chords.}
\label{fig:chords-3}
\end{figure}

Both {\it suspended} and {\it power chords} are similar to triads in that they contain a root and a perfect fifth. They differ, however, in their omission of the third. As shown in Figure~\ref{fig:chords-3}, {\it suspended second chords} (sus2) use a second as replacement for this third, forming a major second (2 half steps) with the root, while {\it suspended fourth chords} (sus4) employ a perfect fourth as replacement \citep{taylor:book89}. The suspended second and fourth often resolve to a more stable third. In addition to these two kinds of suspended chords, our system considers suspended fourth chords that contain an added minor seventh, forming a {\it dominant seventh suspended fourth chord} (7sus4).

In contrast with suspended chords, power chords (pow) do not contain a replacement for the missing third. They simply consist of a root and a perfect fifth. Though they are not formally considered to be chords in classical music, they are commonly referred to in both rock and pop music \citep{denyer:book92}.

\subsection{Chord Ambiguity}
\label{sec:ambiguity}

Sometimes, the same set of notes can have multiple chord interpretations. For example, the  German sixth chord shown in Figure~\ref{fig:chords-2} can also be interpreted as an F dominant seventh chord. Added notes can also lead to other types of ambiguity, for example \{D, F, A, C\} could be an F major sixth chord (i.e. F major with an added sixth) or a D minor seventh chord (i.e. D minor chord with an added minor seventh). Human annotators can determine the correct chord interpretation based on cues such as inversions and context. The semi-CRF model described in this paper captures inversions through the bass features (Appendix~\ref{sec:bass}), whereas context is taken into account through the chord bigram features (Appendix~\ref{sec:chord-bigrams}). This could be further improved by adding other features, such as determining how notes in the current chord resolve to notes in the next chord.

\section{Figuration Heuristics}
\label{sec:appendix-figuration}

\begin{figure}[h]
\centering
\includegraphics[width=\columnwidth]{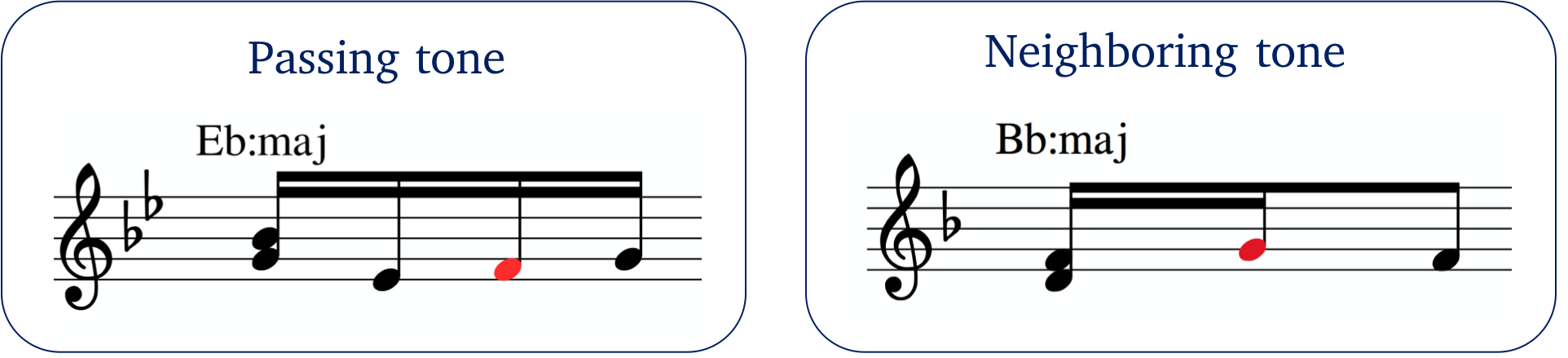}
\caption{Examples of figuration notes, in red.}
\label{fig:passing-neighbor}
\end{figure}

\begin{figure}[h]
\centering
\includegraphics[width=0.9\columnwidth]{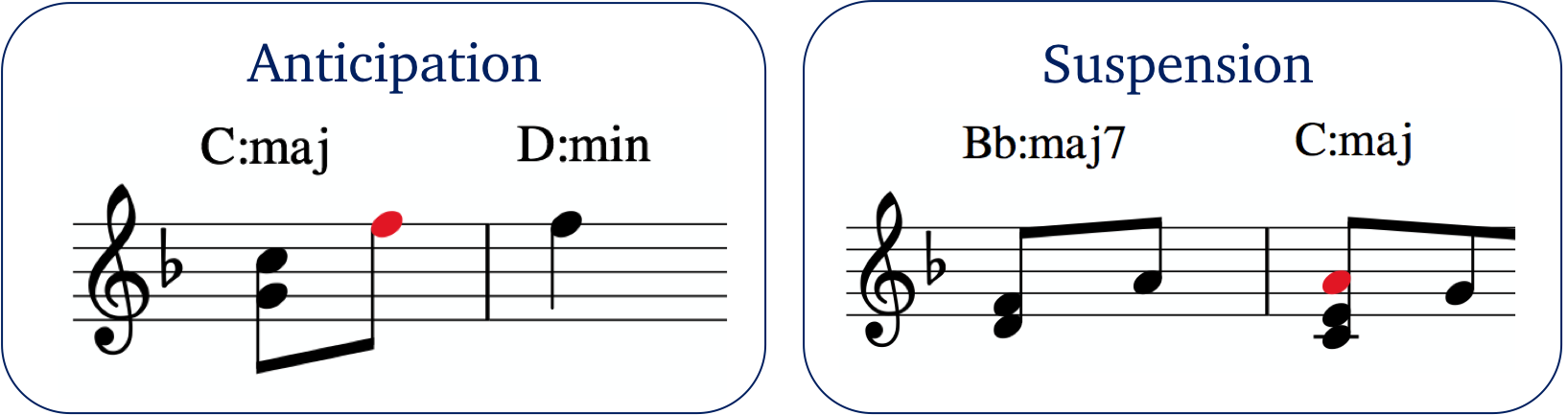}
\caption{Examples of figuration notes, in red.}
\label{fig:suspension-anticipation}
\end{figure}

We designed a set of heuristics to determine whether a note $n$ from a segment $s$ is a figuration note with respect to a candidate chord label $y$. The heuristic rules shown below discover four types of figurations: passing and neighbor notes (Figure~\ref{fig:passing-neighbor}), and suspensions and anticipations (Figure~\ref{fig:suspension-anticipation}).

{\bf Passing}: There are two anchor notes $n_1$ and $n_2$ such that: $n_1$'s offset coincides with $n$'s onset; $n_2$'s onset coincides with $n$'s offset; $n_1$ is one scale step below $n$ and $n_2$ is one step above $n$, or $n_1$ is one step above $n$ and $n_2$ one step below; $n$ is not longer than either $n_1$ or $n_2$; the accent value of $n$ is strictly smaller than the accent value of $n_1$; at least one of the two anchor notes belongs to segment $s$; $n$ is non-harmonic with respect to chord $y$, i.e. $n$ is not equivalent to the root, third, fifth, or added note of $y$; both $n_1$ and $n_2$ are harmonic with respect to the segments they belong to.

{\bf Neighbor}: There are two anchor notes $n_1$ and $n_2$ such that: $n_1$'s offset coincides with $n$'s onset; $n_2$'s onset coincides with $n$'s offset; $n_1$ and $n_2$ are both either one step below or one step above $n$; $n$ is not longer than either $n_1$ or $n_2$; the accent value of $n$ is strictly smaller than the accent value of $n_1$; at least one of the two anchor notes belongs to segment $s$; $n$ is non-harmonic with respect to chord $y$; both anchor notes are harmonic with respect to the segments they belong to.

{\bf Suspension}: Note $n$ belongs to the first event of segment $s$. There is an anchor note $m$ in the previous event (last event in the previous segment) such that: $m$ and $n$ have the same pitch; $n$ is either tied with $m$ (i.e. held over) or $m$'s offset coincides with $n$'s onset (i.e. restruck); $n$ is not longer than $m$; $n$ is non-harmonic with respect to chord $y$, while $m$ is harmonic with respect to the previous chord.

{\bf Anticipation}: Note $n$ belongs to the last event of segment $s$. There is an anchor note $m$ in the next event (first event in the next segment) such that: $n$ and $m$ have the same pitch; $m$ is either tied with $n$ (i.e. held over) or $n$'s offset coincides with $m$'s onset (i.e. restruck); $n$ is not longer than $m$; $n$ is non-harmonic with respect to chord $y$, while $m$ is harmonic relative to all other notes in its event.

Furthermore, because the weak semi-CRF features shown in Equation~\ref{eq:weak2} do not have access to the candidate label $y_{k-1}$ of the previous segment $s_{k-1}$, we need a heuristic to determine whether an anchor note is harmonic whenever the anchor note belongs to the previous segment.
The heuristic simply looks at the other notes in the event containing the anchor note: if the event contains 2 or more other notes, at least 2 of them need to be consonant with the anchor, i.e. intervals of octaves, fifths, thirds, and their inversions; if the event contains just one note other than the anchor note, it has to be consonant with the anchor.

We emphasize that the rules mentioned above for detecting figuration notes are only approximations. We recognize that correctly identifying figuration notes can also depend on subtler stylistic and contextual cues, thus allowing for exceptions to each of these rules.

\section{Chord Recognition Features}
\label{sec:appendix-features}

\newcounter{fi}
\stepcounter{fi}

Given a segment $s$ and chord $y$, we will use the following notation:
\begin{itemize}
  \item $s.\mathit{Notes}$, $s.N$ = the set of notes in the segment $s$.
  \item $s.\mathit{Events}$, $s.E$ = the sequence of events in $s$.
  \item $e.len$, $n.len$ = the length (i.e. duration) of event $e$ or note $n$, in quarters.
  \item $e.acc$, $n.acc$ = the {\it accent value} of event $e$ or note $n$, as computed by the {\tt beatStrength()} function in Music21\footnote{Link to Music21: \url{http://web.mit.edu/music21}}.
  \item $y.\mathit{root}$, $y.\mathit{third}$, and $y.\mathit{fifth}$ = the triad tones of the chord $y$.
  \item $y.\mathit{added}$ = the added note of chord $y$, if $y$ is an added tone chord.
  \item $s.\mathit{Fig}(y)$ = the set of notes in $s$ that are {\it figuration} with respect to chord $y$.
  \item $s.\mathit{NonFig}(y) = s.\mathit{Notes} - s.\mathit{Fig}(y)$ = the set of notes in $s$ that are not figuration with respect to $y$.
\end{itemize}
Note that a note may contain multiple events, as such the note length $n.len$ can be seen as the sum of the length of all events that span the duration of that note. For example, the first G3 in the bass of Figure~\ref{fig:segments2} has a length of a quarter -- it corresponds to the G3 in measure 2 of Figure~\ref{fig:segments} and is shown as a tied note to simplify the description. Therefore its $n.len = 1$. Each of the two events that span its duration have a length of an eighth, hence $e_1.len = e_2.len = 0.5$.

The {\it accent value} is determined based on the metrical position of a note or event, e.g. in a song written in a 4/4 time signature, the first beat position would have a value of 1.0, the third beat 0.5, and the second and fourth beats 0.25. Any other eighth note position within a beat would have a value of 0.125, any sixteenth note position strictly within the beat would have a value of 0.0625, and so on. To determine whether a note $n$ from a segment $s$ is a {\it figuration} note with respect to a candidate chord label $y$, we use a set of heuristics, as detailed in Appendix~\ref{sec:appendix-figuration}.

The duration and accent-weighted segment-level features introduced in this section have real values. Given a real-valued feature $f(s, y)$ that takes values in $[0, 1]$, we discretize it into $K+2$ Boolean features by partitioning the $[0, 1]$ interval into a set of $K$ subinterval bins $\mathcal{B} = \{(b_{k-1}, b_k] | 1 \leq k \leq K\}$. For each bin, the corresponding Boolean feature determines whether $f(s, y) \in (b_{k-1}, b_k]$. Additionally, two Boolean features are defined for the boundary cases $f(s, y) = 0$ and $f(s, y) = 1$. For each real-valued feature, unless specified otherwise, we use the bin set $\mathcal{B} = [0, 0.1, ..., 0.9, 1.0]$.

\subsection{Segment Purity}
\label{sec:purity}
The segment purity feature $f_{\thefi}(s, y)$ computes the fraction of the notes in segment $s$ that are harmonic, i.e. belong to chord $y$:
\begin{equation}
  f_{\thefi}(s, y) = \frac{\displaystyle\sum_{n \in s.\mathit{Notes}} \mathbf{1}[n \in y]}{|s.\mathit{Notes}|} \nonumber
\end{equation}
\stepcounter{fi}
The duration-weighted version $f_{\thefi}(s, y)$ of the purity feature weighs each note $n$ by its length $n.len$:
\begin{equation}
  f_{\thefi}(s, y) = \frac{\displaystyle\sum_{n \in s.\mathit{Notes}} \mathbf{1}[n \in y] * n.len}{\displaystyle\sum_{n \in s.\mathit{Notes}} n.len} \nonumber
\end{equation}
\stepcounter{fi}
The accent-weighted version $f_{\thefi}(s, y)$ of the purity feature weighs each note $n$ by its accent weight $n.acc$:
\begin{equation}
  f_{\thefi}(s, y) = \frac{\displaystyle\sum_{n \in s.\mathit{Notes}} \mathbf{1}[n \in y] * n.acc}{\displaystyle\sum_{n \in s.\mathit{Notes}} n.acc} \nonumber
\end{equation}
\stepcounter{fi}
The 3 real-valued features are discretized using the default bin set $\mathcal{B}$.

\subsubsection{Figuration-Controlled Segment Purity}
\label{sec:purity-fig}

For each segment purity feature, we create a figuration-controlled version that ignores notes that were heuristically detected as figuration, i.e. replace $s.\mathit{Notes}$ with $s.\mathit{NonFig}(y)$ in each feature formula.

\subsection{Chord Coverage}
\label{sec:coverage}

The chord coverage features determine which of the chord notes belong to the segment. In this section, each of the coverage features are non-zero only for major, minor, and diminished triads and their added note counterparts. This is implemented by first defining an indicator function $y.\mathit{Triad}$ that is 1 only for triads and chords with added notes, and then multiplying it into all the triad features from this section.
\begin{eqnarray*}
  y.\mathit{Triad} & = &  \mathbf{1}[y.\mathit{mode}  \in \{\mbox{maj, min, dim}\}]
\end{eqnarray*}
Furthermore, we compress notation by showing the mode predicates as attributes of the label, e.g. $y.maj$ is a predicate equivalent with testing whether $y.\mathit{mode} = \mbox{maj}$. Thus, an equivalent formulation of $y.\mathit{Triad}$ is as follows:
\begin{eqnarray*}
  y.\mathit{Triad} & = &  \mathbf{1}[y.maj \vee y.min \vee y.dim]
\end{eqnarray*}
To avoid clutter, we do not show $y.\mathit{Triad}$ in any of the features below, although it is assumed to be multiplied into all of them. The first 3 coverage features refer to the triad notes:
\begin{eqnarray*}
  f_{\thefi}(s, y) & = & \mathbf{1}[y.\mathit{root} \in s.\mathit{Notes}] \\ \stepcounter{fi}
  f_{\thefi}(s, y) & = & \mathbf{1}[y.\mathit{third} \in s.\mathit{Notes}] \\ \stepcounter{fi}
  f_{\thefi}(s, y) & = & \mathbf{1}[y.\mathit{fifth} \in s.\mathit{Notes}] \stepcounter{fi}
\end{eqnarray*}
A separate feature determines if the segment contains all the notes in the chord:
\begin{equation}
  f_{\thefi}(s, y) = \displaystyle\prod_{n \in y} \mathbf{1}[n \in s.\mathit{Notes}] \nonumber
\end{equation}
\stepcounter{fi}
A chord may have an added tone $y.\mathit{added}$, such as a 4th, a 6th, or a 7th. If a chord has an added tone, we define two features that determine whether the segment contains the added note:
\begin{eqnarray*}
  f_{\thefi}(s, y) & = & \mathbf{1}[\exists y.\mathit{added} \wedge y.\mathit{added} \in s.\mathit{Notes}] \\ \stepcounter{fi}
  f_{\thefi}(s, y) & = & \mathbf{1}[\exists y.\mathit{added} \wedge y.\mathit{added} \notin s.\mathit{Notes}] \stepcounter{fi}
\end{eqnarray*}
Through the first feature, the system can learn to prefer the added tone version of the chord when the segment contains it, while the second feature enables the system to learn to prefer the triad-only version if no added tone is in the segment. To prevent the system from recognizing added chords too liberally, we add a feature that is triggered whenever the total length of the added notes in the segment is greater than the total length of the root:
\begin{eqnarray*}
  alen(s, y) = \displaystyle\sum_{n \in s.\mathit{Notes}} \!\!\!\! \mathbf{1}[n = y.\mathit{added}] * n.len \\
  rlen(s, y) = \displaystyle\sum_{n \in s.\mathit{Notes}} \!\!\!\! \mathbf{1}[n = y.\mathit{root}] * n.len \\
  f_{\thefi}(s, y) = \mathbf{1}[\exists y.\mathit{added}] * \mathbf{1}[alen(s, y) > rlen(s, y)]
\end{eqnarray*}

\stepcounter{fi}
The duration-weighted versions of the chord coverage features weigh each chord tone by its total duration in the segment. For the root, the feature would be computed as shown below:
\begin{equation}
  f_{\thefi}(s, y) = \frac{\displaystyle\sum_{n \in s.\mathit{Notes}} \mathbf{1}[n = y.\mathit{root}] * n.len}{\displaystyle\sum_{n \in s.\mathit{Notes}} n.len} \nonumber
\end{equation}
\stepcounter{fi}
Similar features $f_{\thefi}$\stepcounter{fi} and $f_{\thefi}$\stepcounter{fi} are computed for the third and the fifth. The corresponding accent-weighted features $f_{\thefi}$\stepcounter{fi}, $f_{\thefi}$\stepcounter{fi}, and $f_{\thefi}$\stepcounter{fi} are computed in a similar way, by replacing the note duration $n.len$ in the duration-weighted formulas with the note accent value $n.acc$.

The duration-weighted feature for the added tone is computed similarly:
\begin{equation}
  f_{\thefi}(s, y) = \frac{\mathbf{1}[\exists y.\mathit{added}] * \!\!\!\!\displaystyle\sum_{n \in s.\mathit{Notes}} \!\!\!\! \mathbf{1}[n = y.\mathit{added}] * n.len}{\displaystyle\sum_{n \in s.\mathit{Notes}} n.len} \nonumber
\end{equation}
\stepcounter{fi}
Furthermore, by replacing $n.len$ with $n.acc$, we also obtain the accent-weighted version $f_{\thefi}$\stepcounter{fi}.

An alternative definition of duration-weighted features is based on the proportion of the segment time that is covered by a particular chord note. The corresponding duration-weighted feature for the chord root is shown below:
\begin{equation}
  f_{\thefi}(s, y) = \frac{\displaystyle\sum_{e \in s.\mathit{Events}} \mathbf{1}[y.\mathit{root} \in e] * e.len}{\displaystyle\sum_{e \in s.\mathit{Events}} e.len} \nonumber
\end{equation}
\stepcounter{fi}
Similar duration-weighted features normalized by the segment length are defined for thirds, fifths, and added notes.

All duration-weighted and accent-weighted features are discretized using the default bin set $\mathcal{B}$.

\subsubsection{Chord Coverage for Augmented 6th Chords}
\label{sec:as-coverage}

\newcounter{asi}
\stepcounter{asi}

We label each note appearing in an augmented 6th chord as follows:
\begin{itemize}
  \item $y.bass$ = the lowest note.
  \item $y.3rd$ = the note that is a third above $y.bass$.
  \item $y.6th$ = the note that is an augmented sixth above $y.bass$.
  \item $y.5th$ = the note that forms a perfect fifth (for German 6th chords) or a diminished fifth (for French) above $y.bass$.
\end{itemize}
The features defined in this section are non-zero only for augmented 6th chord labels. Similar to Appendix~\ref{sec:coverage}, we define an indicator function $y.\mathit{AS}$ that is 1 only for augmented 6th chords and implicitly multiply this into each of the features from this section.
\begin{eqnarray*}
  y.\mathit{AS} & = &  \mathbf{1}[y.\mathit{mode}  \in \{\mbox{it6, fr6, ger6}\}] \\
  y.\mathit{AS} & = &  \mathbf{1}[y.it6 \vee y.fr6 \vee y.ger6]
\end{eqnarray*}
We define an additional indicator function $y.\mathit{FG}$ that is 1 only for French and German 6th chords.
\begin{eqnarray*}
  y.\mathit{FG} & = &  \mathbf{1}[y.\mathit{fr6} \vee y.\mathit{ger6}]
\end{eqnarray*}
The coverage features for augmented 6th chords are overall analogous to the ones for triad chords.
\begin{eqnarray*}
  as_{\theasi}(s, y) & = & \mathbf{1}[y.\mathit{bass} \in s.\mathit{Notes}] \\ \stepcounter{asi}
  as_{\theasi}(s, y) & = & \mathbf{1}[y.\mathit{3rd} \in s.\mathit{Notes}] \\ \stepcounter{asi}
  as_{\theasi}(s, y) & = & \mathbf{1}[y.\mathit{6th} \in s.\mathit{Notes}] \\ \stepcounter{asi}
  as_{\theasi}(s, y) & = & \mathbf{1}[y.FG \wedge y.\mathit{5th} \in s.\mathit{Notes}] \stepcounter{asi}
  \end{eqnarray*} 
 The duration-weighted versions are as follows:
  \begin{equation}
  as_{\theasi}(s, y) = \frac{\displaystyle\sum_{n \in s.\mathit{Notes}} \mathbf{1}[n = y.\mathit{bass}] * n.len}{\displaystyle\sum_{n \in s.\mathit{Notes}} n.len} \nonumber \stepcounter{asi}
  \end{equation}

  \begin{equation}
  as_{\theasi}(s, y) = \mathbf{1}[y.FG] * \frac{\displaystyle\sum_{n \in s.\mathit{Notes}} \mathbf{1}[n = y.\mathit{5th}] * n.len}{\displaystyle\sum_{n \in s.\mathit{Notes}} n.len} \nonumber \\
\end{equation}
As before, we replace $n.len$ with $n.acc$ to obtain the accent-weighted versions of $as_{5}$ and $as_{\theasi}$\stepcounter{asi}. We also define segment-based duration-weighted features:
  \begin{equation}
  as_{\theasi}(s, y) = \frac{\displaystyle\sum_{e \in s.\mathit{Events}} \mathbf{1}[y.\mathit{bass} \in e] * e.len}{\displaystyle\sum_{e \in s.\mathit{Events}} e.len} \nonumber \stepcounter{asi}
\end{equation}

\subsubsection{Chord Coverage for Suspended and Power Chords}
\label{sec:sp-coverage}

\newcounter{spi}
\stepcounter{spi}

As before, we define the features in this section to be non-zero only for suspended or power chord labels by implicitly multiplying them with an indicator function $y.\mathit{SP}$.
\begin{eqnarray*}
  y.\mathit{SP} & = &  \mathbf{1}[y.sus2 \vee y.sus4 \vee y.7sus4 \vee y.pow]
\end{eqnarray*}
The coverage features for suspended and power chords are also similar to the ones defined for triad chords.
\begin{eqnarray*}
  sp_{\thespi}(s, y) & = & \mathbf{1}[y.\mathit{root} \in s.\mathit{Notes}] \\ \stepcounter{spi}
  sp_{\thespi}(s, y) & = & \mathbf{1}[y.sus2 \wedge y.\mathit{2nd} \in s.\mathit{Notes} \vee \\ && \ \ (y.sus4 \vee y.7sus4) \wedge y.\mathit{4th} \in s.\mathit{Notes})] \\ \stepcounter{spi}
  sp_{\thespi}(s, y) & = & \mathbf{1}[y.\mathit{5th} \in s.\mathit{Notes}] \\ \stepcounter{spi}
  sp_{\thespi}(s, y) & = & \mathbf{1}[y.7sus4 \wedge y.\mathit{7th} \in s.\mathit{Notes}] \\ \stepcounter{spi}
  sp_{\thespi}(s, y) & = & \mathbf{1}[y.7sus4 \wedge y.\mathit{7th} \notin s.\mathit{Notes}] \stepcounter{spi}
  \end{eqnarray*}
  \begin{eqnarray*}
  alen(s, y) & = & \displaystyle\sum_{n \in s.\mathit{Notes}} \!\!\!\! \mathbf{1}[n = y.\mathit{7th}] * n.len \\
  rlen(s, y) & = & \displaystyle\sum_{n \in s.\mathit{Notes}} \!\!\!\! \mathbf{1}[n = y.\mathit{root}] * n.len \\
  sp_{\thespi}(s, y) & = & \mathbf{1}[y.7sus4 \wedge alen(s, y) > rlen(s, y)] \stepcounter{spi}
\end{eqnarray*}
The duration-weighted versions are as follows:
\begin{eqnarray*}
  sp_{\thespi}(s, y) & = &\frac{rlen(s, y)}{\displaystyle\sum_{n \in s.\mathit{Notes}} n.len} \nonumber \stepcounter{spi} \\
  sp_{\thespi}(s, y) & = & \mathbf{1}[y.7sus4] * \frac{alen(s, y)}{\displaystyle\sum_{n \in s.\mathit{Notes}} n.len} \nonumber\stepcounter{spi} \\
\end{eqnarray*}
We also define accent-weighted versions of $sp_7$ and $sp_8$, as well as segment-based duration-weighted features:
\begin{eqnarray*}
  sp_{\thespi}(s, y) & = & \frac{\displaystyle\sum_{e \in s.\mathit{Events}} \mathbf{1}[y.\mathit{root} \in e] * e.len}{\displaystyle\sum_{e \in s.\mathit{Events}} e.len} \nonumber \stepcounter{spi}
\end{eqnarray*}

\subsubsection{Figuration-Controlled Chord Coverage}

For each chord coverage feature, we create a figuration-controlled version that ignores notes that were heuristically detected as figuration, i.e. replace $s.\mathit{Notes}$ with $s.\mathit{NonFig}(y)$ in each feature formula.

\subsection{Bass}
\label{sec:bass}

The bass note provides the foundation for the harmony of a musical segment. For a correct segment, its bass note often matches the root of its chord label. If the bass note instead matches the chord's third or fifth, or is an added dissonance, this may indicate that the chord is inverted. Thus, comparing the bass note with the chord tones can provide useful features for determining whether a segment is compatible with a chord label. As in Appendix~\ref{sec:coverage}, we implicitly multiply each of these features with $y.\mathit{Triad}$ so that they are non-zero only for triads and chords with added notes.

There are multiple ways to define the bass note of a segment $s$. One possible definition is the lowest note of the first event in the segment, i.e. $s.e_1.\mathit{bass}$. Comparing it with the root, third, fifth, and added tones of a chord results in the following features:
\begin{eqnarray*}
  f_{\thefi}(s, y) & = & \mathbf{1}[s.e_1.\mathit{bass} = y.\mathit{root}] \stepcounter{fi}\\
  f_{\thefi}(s, y) & = & \mathbf{1}[s.e_1.\mathit{bass} = y.\mathit{third}] \stepcounter{fi}\\
  f_{\thefi}(s, y) & = & \mathbf{1}[s.e_1.\mathit{bass} = y.\mathit{fifth}] \stepcounter{fi}\\
  f_{\thefi}(s, y) & = & \mathbf{1}[\exists y.\mathit{added} \wedge s.e_1.\mathit{bass} = y.\mathit{added}] \stepcounter{fi}
\end{eqnarray*}
An alternative definition of the bass note of a segment is the lowest note in the entire segment, i.e. $\min_{e \in s.E} e.\mathit{bass}$. The corresponding features will be:
\begin{eqnarray*}
  f_{\thefi}(s, y) & = & \mathbf{1}[y.\mathit{root} = \min_{e \in s.E} e.\mathit{bass}] \stepcounter{fi} \\
  f_{\thefi}(s, y) & = & \mathbf{1}[y.\mathit{third} = \min_{e \in s.E} e.\mathit{bass}] \stepcounter{fi} \\
  f_{\thefi}(s, y) & = & \mathbf{1}[y.\mathit{fifth} = \min_{e \in s.E} e.\mathit{bass}] \stepcounter{fi} \\
  f_{\thefi}(s, y) & = & \mathbf{1}[\exists y.\mathit{added} \wedge y.\mathit{added} = \min_{e \in s.E} e.\mathit{bass}]
\end{eqnarray*}

\stepcounter{fi}
The duration-weighted version of the bass features weigh each chord tone by the time it is used as the lowest note in each segment event, normalized by the duration of the bass notes in all the events. For the root, the feature is computed as shown below:
\begin{equation}
  f_{\thefi}(s, y) = \frac{\displaystyle\sum_{e \in s.\mathit{Events}} \mathbf{1}[e.\mathit{bass} = y.\mathit{root}] * e.len}{\displaystyle\sum_{e \in s.\mathit{Events}} e.len} \nonumber
\end{equation}
\stepcounter{fi}
Similar features $f_{\thefi}$\stepcounter{fi} and $f_{\thefi}$\stepcounter{fi} are computed for the third and the fifth. The duration-weighted feature for the added tone is computed as follows:
\begin{equation}
  f_{\thefi}(s, y) = \frac{\mathbf{1}[\exists y.\mathit{added}] * \!\!\!\!\displaystyle\sum_{e \in s.\mathit{E}} \mathbf{1}[e.\mathit{bass} = y.\mathit{added}] * e.len}{\displaystyle\sum_{e \in s.\mathit{E}} e.len} \nonumber
\end{equation}
\stepcounter{fi}
The corresponding accent-weighted features $f_{\thefi}$\stepcounter{fi}, $f_{\thefi}$\stepcounter{fi}, $f_{\thefi}$\stepcounter{fi}, and $f_{\thefi}$\stepcounter{fi} are computed in a similar way, by replacing the bass duration $e.\mathit{bass}.len$ in the duration-weighted formulas with the note accent value $e.\mathit{bass}.acc$.

All duration-weighted and accent-weighted features are discretized using the default bin set $\mathcal{B}$.

\subsubsection{Bass Features for Augmented 6th Chords}
Similar to the chord coverage features in Appendix~\ref{sec:as-coverage}, we assume that the indicator $y.\mathit{AS}$ is multiplied into all features in this section, which means they are non-zero only for augmented 6th chords.
\begin{eqnarray*}
  as_{\theasi}(s, y) & = & \mathbf{1}[s.e_1.\mathit{bass} = y.\mathit{bass}] \stepcounter{asi}\\
  as_{\theasi}(s, y) & = & \mathbf{1}[s.e_1.\mathit{bass} = y.\mathit{3rd}] \stepcounter{asi}\\
  as_{\theasi}(s, y) & = & \mathbf{1}[s.e_1.\mathit{bass} = y.\mathit{6th}] \stepcounter{asi}\\
  as_{\theasi}(s, y) & = & \mathbf{1}[(y.fr6 \vee y.ger6) \wedge s.e_1.\mathit{bass} = y.\mathit{5th}] \stepcounter{asi}
\end{eqnarray*}
\begin{eqnarray*}
  as_{\theasi}(s, y) & = & \mathbf{1}[y.\mathit{bass} = \min_{e \in s.E} e.\mathit{bass}] \stepcounter{asi} \\
  as_{\theasi}(s, y) & = & \mathbf{1}[y.\mathit{3rd} = \min_{e \in s.E} e.\mathit{bass}] \stepcounter{asi} \\
  as_{\theasi}(s, y) & = & \mathbf{1}[y.\mathit{6th} = \min_{e \in s.E} e.\mathit{bass}] \stepcounter{asi} \\
  as_{\theasi}(s, y) & = & \mathbf{1}[y.FG \wedge y.\mathit{5th} = \min_{e \in s.E} e.\mathit{bass}] \stepcounter{asi}
\end{eqnarray*}
We define the following duration-weighted version for the augmented sixth bass and fifth.
\begin{eqnarray*}
  as_{\theasi}(s, y) & = & \frac{\displaystyle\sum_{e \in s.\mathit{E}} \mathbf{1}[e.\mathit{bass} = y.\mathit{bass}] * e.len}{\displaystyle\sum_{e \in s.\mathit{E}} e.len} \nonumber\stepcounter{asi} \\
  as_{\theasi}(s, y) & = & \mathbf{1}[y.FG] * \frac{\displaystyle\sum_{e \in s.\mathit{E}} \mathbf{1}[e.\mathit{bass} = y.\mathit{5th}] * e.len}{\displaystyle\sum_{e \in s.\mathit{E}} e.len} \nonumber\stepcounter{asi}
\end{eqnarray*}

\subsubsection{Bass Features for Suspended and Power Chords}
\label{sec:sp-bass}

The indicator $y.\mathit{SP}$ is multiplied into all features in this section like in Appendix~\ref{sec:sp-coverage}, meaning they are non-zero only for suspended and power chords.
\begin{eqnarray*}
  sp_{\thespi}(s, y) & = & \mathbf{1}[s.e_1.\mathit{bass} = y.\mathit{root}] \stepcounter{spi}\\
  sp_{\thespi}(s, y) & = & \mathbf{1}[y.sus2 \wedge s.e_1.\mathit{bass} = y.\mathit{2nd} \vee \\ && \ \ (y.sus4 \vee y.7sus4) \wedge s.e_1.\mathit{bass} = y.\mathit{4th}] \stepcounter{spi}\\
  sp_{\thespi}(s, y) & = & \mathbf{1}[s.e_1.\mathit{bass} = y.\mathit{5th}] \stepcounter{spi}\\
  sp_{\thespi}(s, y) & = & \mathbf{1}[y.7sus4 \wedge s.e_1.\mathit{bass} = y.\mathit{7th}] \stepcounter{spi}
\end{eqnarray*}
\begin{eqnarray*}
  sp_{\thespi}(s, y) & = & \mathbf{1}[y.\mathit{root} = \min_{e \in s.E} e.\mathit{bass}] \stepcounter{spi} \\
  sp_{\thespi}(s, y) & = & \mathbf{1}[y.sus2 \wedge y.\mathit{2nd} = \min_{e \in s.E} e.\mathit{bass} \vee \\ && \ \ (y.sus4 \vee y.7sus4) \wedge y.\mathit{4th} = \min_{e \in s.E} e.\mathit{bass}] \stepcounter{spi} \\
  sp_{\thespi}(s, y) & = & \mathbf{1}[y.\mathit{5th} = \min_{e \in s.E} e.\mathit{bass}] \stepcounter{spi} \\
  sp_{\thespi}(s, y) & = & \mathbf{1}[y.7sus4 \wedge y.\mathit{7th} = \min_{e \in s.E} e.\mathit{bass}] \stepcounter{spi}
\end{eqnarray*}
The duration-weighted version for the root and seventh are computed as follows:
\begin{eqnarray*}
  sp_{\thespi}(s, y) \!\!\!\!\!\! & = & \!\!\!\!\!\! \frac{\displaystyle\sum_{e \in s.\mathit{E}} \mathbf{1}[e.\mathit{bass} = y.\mathit{root}] * e.len}{\displaystyle\sum_{e \in s.\mathit{E}} e.len} \nonumber\stepcounter{spi} \\
  sp_{\thespi}(s, y) \!\!\!\!\!\! & = & \!\!\!\!\!\! \mathbf{1}[y.7sus4] * \frac{\displaystyle\sum_{e \in s.\mathit{E}} \mathbf{1}[e.\mathit{bass} = y.\mathit{7th}] * e.len}{\displaystyle\sum_{e \in s.\mathit{E}} e.len} \nonumber
\end{eqnarray*}

\subsubsection{Figuration-Controlled Bass}

For each bass feature, we create a figuration-controlled version that ignores event bass notes that were heuristically detected as figuration, i.e. replace $e \in s.\mathit{Events}$ with $e \in s.\mathit{Events} \wedge e.bass \notin s.\mathit{Fig}(y)$ in each feature formula.

\subsection{Chord Bigrams}
\label{sec:chord-bigrams}

\newcounter{gi}
\stepcounter{gi}

The arrangement of chords in chord progressions is an important component of {\it harmonic syntax} \citep{aldwell:book11}. A first-order semi-Markov CRF model can capture chord sequencing information only through the chord labels $y$ and $y'$ of the current and previous segment. To obtain features that generalize to unseen chord sequences, we follow \citet{radicioni:amir10} and create chord bigram features using only the {\it mode}, the {\it added} note, and the interval in semitones between the roots of the two chords. We define the possible modes of a chord label as follows:
\begin{eqnarray*}
  \mathcal{M} & = & \{\mbox{maj, min, dim}\} \nonumber \\
              &   & \hspace{1em} \cup \; \{\mbox{it6, fr6, ger6}\} \nonumber \\
              &   & \hspace{2em} \cup \; \{\mbox{sus2, sus4, 7sus4, pow}\} \nonumber
\end{eqnarray*}
Other than the common major (maj), minor (min), and diminished (dim) modes, the following chord types have been included in $\mathcal{M}$ as modes:
\begin{itemize}
  \item Augmented 6th chords: Italian 6th (it6), French 6th (fr6), and German 6th (ger6).
  \item Suspended chords: suspended second (sus2), suspended fourth (sus4), dominant seventh suspended fourth (7sus4).
  \item Power (pow) chords.
\end{itemize}
Correspondingly, the chord bigrams can be generated using the feature template below:
\begin{eqnarray*}
  g_{\thegi}(y, y') & =
      & \mathbf{1}[(y.\mathit{mode}, y'.\mathit{mode}) \in \mathcal{M} \times \mathcal{M} \nonumber \\ 
   &  &  \wedge (y.\mathit{added}, y'.\mathit{added}) \in \{\emptyset, 4, 6, 7\} \times \{\emptyset, 4, 6, 7\} \nonumber \\ 
   &  &  \wedge \left|y.\mathit{root} - y'.\mathit{root}\right| = \{0, 1, ..., 11\}] \nonumber
\end{eqnarray*}
Note that $y.root$ is replaced with $y.bass$ for augmented 6th chords. Additionally, $y.added$ is always none ($\emptyset$) for augmented 6th, suspended, and power chords. Thus, $g_{\thegi}(y, y')$ is a feature template that can generate (3 triad modes $\times$ 4 added + 3 aug6 modes + 3 sus modes + 1 pow mode)$^2 \times$ 12 intervals = 4,332 distinct features. To reduce the number of features, we use only the {\it (mode.added)--(mode.added)'--interval} combinations that appear in the manually annotated chord bigrams from the training data.

\stepcounter{gi}

\subsection{Chord Changes and Metrical Accent}
\label{sec:chord-accent}

In general, repeating a chord creates very little accent, whereas changing a chord tends to attract an accent \citep{aldwell:book11}. Although conflict between meter and harmony is an important compositional resource, in general chord changes support the meter. Correspondingly, a new feature is defined as the accent value of the first event in a candidate segment:
\begin{equation}
  f_{\thefi}(s, y) = s.e_1.acc \nonumber \stepcounter{fi}
\end{equation}

%
%
%
%

\end{document}